\documentclass[twocolumn]{aastex631}

\usepackage{orcidlink}
\usepackage{bm}

\usepackage{breqn}
\usepackage{graphicx}	

\def\m{\ensuremath\mathbf}

\begin{document}

\title{Statistical recovery of 21cm visibilities and their power spectra with
Gaussian constrained \\
realisations and Gibbs sampling}

\author{Fraser~Kennedy\,\orcidlink{0000-0002-5883-6543}\!}
\affiliation{Astronomy Unit, Queen Mary University of London, Mile End Road, London E1 4NS, United Kingdom}

\author{Philip~Bull\,\orcidlink{0000-0001-5668-3101}\!}
\affiliation{Jodrell Bank Centre for Astrophysics, University of Manchester, Manchester, M13 9PL, United Kingdom}
\affiliation{Department of Physics and Astronomy, University of Western Cape, Cape Town 7535, South Africa}

\author{Michael~J.~Wilensky\,\orcidlink{0000-0001-7716-9312}\!}
\affiliation{Jodrell Bank Centre for Astrophysics, University of Manchester, Manchester, M13 9PL, United Kingdom}

\author{Jacob~Burba\,\orcidlink{0000-0002-8465-9341}\!}
\affiliation{Jodrell Bank Centre for Astrophysics, University of Manchester, Manchester, M13 9PL, United Kingdom}

\author{Samir~Choudhuri\,\orcidlink{0000-0002-2338-935X}\!}
\affiliation{Centre for Strings, Gravitation and Cosmology, Department of Physics, Indian Institute of Technology Madras, Chennai 600036, India}

\begin{abstract}
Radio interferometers designed to probe the 21cm signal from Cosmic Dawn and the Epoch of Reionisation must contend with systematic effects that make it difficult to achieve sufficient dynamic range to separate the 21cm signal from foreground emission and other effects. For instance, the instrument's chromatic response modulates the otherwise spectrally smooth foregrounds, making them difficult to model, while a significant fraction of the data must be excised due to the presence of radio frequency interference (RFI), leaving gaps in the data. Errors in modelling the (modulated and gappy) foregrounds can easily generate spurious contamination of what should otherwise be 21cm signal-dominated modes. Various approaches have been developed to mitigate these issues by (e.g.) using non-parametric reconstruction of the foregrounds, in-painting the gaps, and weighting the data to reduce the level of contamination.
We present a Bayesian statistical method that combines these approaches, using the coupled techniques of Gaussian constrained realisations (GCR) and Gibbs sampling. This provides a way of drawing samples from the joint posterior distribution of the 21cm signal modes and their power spectrum in the presence of gappy data and an uncertain foreground model in a computationally scalable manner. The data are weighted by an inverse covariance matrix that is estimated as part of the inference, along with a foreground model that can then be marginalised over. We demonstrate the application of this technique on a simulated HERA-like delay spectrum analysis, comparing three different approaches for accounting for the foreground components. 
\end{abstract}

\keywords{dark ages, reionization, first stars --- methods: statistical ---  techniques: interferometric}

\section{Introduction} \label{sec:intro}

The redshifted 21cm emission line from neutral hydrogen is much-anticipated as a sensitive probe of the thermal history of the first billion years or so of cosmic time, particularly the periods when the first stars and galaxies formed (Cosmic Dawn) and then reionised the intergalactic medium (the Epoch of Reionisation, or EoR) \citep{1997ApJ...475..429M, 2006PhR...433..181F, morales10, 2012RPPh...75h6901P}. A number of radio interferometers have been designed specifically to measure the statistical fluctuations in this signal, which arises in the approximate redshift range $6 \lesssim z \lesssim 27$, corresponding to frequencies of $\sim 50-200$~MHz. Examples of those searching for the signal from the EoR include the Giant Meterwave Radio Telescope \citep[GMRT; ][]{swarup91,Paciga2013}; the Murchison Wide-field Array \citep[MWA;][]{tingay13,2018PASA...35...33W, trott20}; the Low Frequency Array \citep[LOFAR;][]{vanhaarlem13, 2017ApJ...838...65P, 2020MNRAS.493.1662M}; the Long Wavelength Array \citep[LWA;][]{2014era..conf10203H, 2021MNRAS.506.5802G}; the Precision Array to Probe the Epoch of Reionization \citep[PAPER;][]{2010AJ....139.1468P, kolopanis2019}; and the Hydrogen Epoch of Reionization Array \citep[HERA;][]{2017PASP..129d5001D, 2022ApJ...925..221A}. The usual goal of these experiments is to build up the large sensitivity required to make a statistical detection of the power spectrum of the brightness temperature fluctuations of the 21cm line as a function of redshift, by observing for long periods of time with large numbers of receiving elements and baselines. The presence of bright foreground emission places extremely stringent requirements on the fidelity of the instrumental calibration required to detect the 21cm signal however \citep{Barry:2016cpg, 2016MNRAS.463.4317P, 2017MNRAS.470.1849E, 2018MNRAS.478.1484G, joseph18, byrne19, 2019A&A...622A...5D, 2019MNRAS.483.5480M}, with other systematic effects such as mode mixing \citep[e.g.][]{2012ApJ...752..137M}, non-redundancy \citep[e.g.][]{Orosz:2018avj, joseph20, 2021MNRAS.506.2066C}, in-painting artifacts \citep[e.g.][]{2022ApJ...929..104C, Pagano2022}, reflection and coupling artifacts \citep[e.g.][]{2016MNRAS.460.4320E, 2019ApJ...884..105K} as well as polarization leakage and the ionosphere \citep{2018MNRAS.478.1484G, 2022arXiv220712090K} needing to be handled carefully.

The foreground emission is particularly challenging because of the high dynamic range between it and the target 21cm signal, which is expected to be somewhere in the region of $10^4-10^5$ times fainter (in temperature). This in itself is not too problematic if the foreground signal is well-segregated from the 21cm signal when projected onto a suitable basis, such that the foreground emission can be localised to a handful of modes that can then be modelled or filtered out. Since the dominant source of foreground emission is synchrotron radiation, which has a smooth (power-law) frequency spectrum, an effective localisation of the foregrounds into only the smoothest components of a harmonic/Fourier basis could reasonably be achieved \citep{2002ApJ...564..576D, santos05, ali08}. This is unfortunately disrupted by the complicated spectral response of the instruments themselves, which modulate the sky signal, introducing additional spectral features that are generally difficult to model with sufficiently high precision. As a result, the instrumental response (and errors in its calibration) `scatters' foreground emission outside of its intrinsic localisation region, with even a small amount of scattering (say, at the $10^{-4}$ level) able to swamp the 21cm emission. Even with precise instrumental calibration, a large wedge-shaped region of foreground contamination is introduced into the 2D Fourier space of the interferometric visibilities by instrumental effects \citep{2010ApJ...724..526D, 2012ApJ...752..137M, 2013ApJ...768L..36P, 2015ApJ...804...14T}, and must either be filtered out \citep[`foreground avoidance'; e.g.][]{2014PhRvD..90b3018L, 2014PhRvD..90b3019L, 2015ApJ...804...14T} or modelled and subtracted \citep[`foreground mitigation'; e.g.][]{2002ApJ...564..576D, santos05, 2006ApJ...648..767M, 2009ApJ...695..183B, 10.1093/mnras/stw161}.

The difficulty of handling the foreground contamination makes it especially important to avoid analysis steps that could introduce further scattering of the foregrounds, e.g. through coupling of Fourier modes inside and outside the wedge. Unfortunately, a number of other systematic effects can naturally cause scattering of this kind. Spurious artificial radio emission (radio frequency interference, or RFI) poses a particular challenge for low frequency experiments, because many transmitters operate inside the experimental band, leaving extremely bright signals in (typically narrow) regions in frequency that effectively render that part of the data unrecoverable during the transmitter's time of operation. These regions must be aggressively identified and masked to prevent substantial contamination of the data \citep[e.g. see][]{2019ApJ...884....1B, 2019ApJ...887..141L, 2020MNRAS.493.1662M, 2020MNRAS.498..265W}. The mask itself then becomes the issue; masking introduces discontinuous jumps in the data, which are problematic for steps of the analysis that rely on harmonic transforms, such as the Fourier transforms that are used during power spectrum estimation \citep{2019MNRAS.484.2866O, 2022MNRAS.510.5023W}. When a Fourier transform is applied to a step or spike, ringing artifacts are generated that strongly couple modes inside the foreground wedge to those outside, causing widespread leakage of the foreground emission across the rest of the Fourier space. 

Methods must therefore be found that can prevent or suppress the ringing in order to successfully measure the power spectrum. These typically fall into two camps: {\it in-painting}, which replaces the missing/masked data with a plausible model \citep[e.g. CLEAN, DAYENU, and Gaussian Process Regression;][]{2009AJ....138..219P, 2019MNRAS.484.2866O, 2020MNRAS.493.1662M, 2021MNRAS.500.5195E, 2021MNRAS.501.1463K, 2022ApJ...929..104C}; and harmonic analysis methods \citep[e.g. LSSA;][]{2016ApJ...818..139T, 2017ApJ...838...65P, 2022ApJ...929..104C} that account for non-uniformly sampled data. The general goal of these methods is to allow the 21cm power spectrum to be recovered in an unbiased manner, without ringing artifacts. Beyond this, other desirable features may be sought, such as minimising the variance of the power spectrum estimates, avoiding strong model dependence, interpretability of the power spectrum estimate, computational efficiency, or correct propagation of uncertainty.

As a final complication, we would also like to apply an inverse covariance weighting to the data in order to recover an optimal power spectrum estimate while also accounting for correlations and mode mixing. This requires a high-fidelity estimate of the true covariance matrix of the data, which is generally not available. Due to the inherently high dynamic range, the signal eigenmodes in the data covariance matrix have small eigenvalues, and are easily mis-estimated. Using the empirical covariance matrix measured from the data as an estimate leads to signal loss in the quadratic estimator formalism, since the estimated signal becomes a quartic (rather than quadratic) function of the data \citep{kolopanis2019}. Simulations rely on empirical sky models that are incomplete.

Solving these problems requires a method that is able to account properly for missing frequency space data, model the foregrounds (as corrupted by the instrument), and weight the data in an optimal way.
In this paper, we present a Bayesian signal recovery method that aims to capture the statistical interactions between the components of the data in order to achieve these aims.  
Our goal is to estimate the joint posterior distribution of a model that is sufficiently flexible to recover the 21cm signal and foregrounds without requiring strong model assumptions. From the joint posterior, we can then derive best-fit models of the components, their uncertainties, and any correlations between them.

Flexible models typically require large numbers of parameters. Our method is based on Gibbs sampling \citep{geman1984stochastic}, which provides a way of sampling from the joint posterior distribution of a model with many parameters by iteratively sampling from a set of more tractable conditional distributions instead. This has been used to good effect in cosmic microwave background (CMB) inference problems such as foreground separation \citep{2004PhRvD..70h3511W, Eriksen}, and has also recently been applied to 21cm data at lower redshift to estimate power spectra in the presence of masked data \citep{2022arXiv220201242C}. We follow the structure of the method in \citet{Eriksen}, which used a Gibbs sampling scheme to recover the joint posterior distribution of the CMB signal field and covariance, as well as various foreground parameters. 
In our case, we aim to recover the joint posterior distribution of the (baseline dependent) EoR 21cm signal visibilities, their power spectra, and a foreground model in the presence of missing frequency channels in the data. We compare three Gibbs sampling implementations designed to achieve that goal:
\begin{itemize}
 \item A `total signal' sampler, which does not differentiate between the EoR signal and the foregrounds, aiming to produce an estimate of the total EoR plus foreground delay spectrum;
 \item A sampler which models foregrounds using the eigenmodes of simulated foreground covariance matrices;
 \item A sampler which jointly samples the foregrounds along with the signal, conditioning on the foreground covariance matrix being known.
 \end{itemize}
We apply these methods to simulations of visibility data with realistic point source foregrounds, a simple Gaussian EoR signal model, and various patterns of gaps to mimic the effect of RFI flagging. We test these models under different noise conditions, and also examine their robustness to incomplete sky models used to generate the priors/models for the foreground component.

The structure of the paper is as follows. In Section~\ref{sec:bayesianrecovery} we provide some grounding in Bayesian methods relevant to Gibbs sampling, namely Wiener filtering and the Gaussian constrained realistion equation, and detail three examples of Gibbs sampling implementations that can be used to recover the EoR signal. In Section~\ref{sec:sims} we describe the visibility simulations we use in the rest of our study. In Section~\ref{sec:results} we present the performance of each of the Gibbs sampling implementations on our simulated data under different flagging and noise conditions, and compare the achieved recovery in each case. We conclude in Section~\ref{sec:conclusions}.

\section{Bayesian recovery of the 21cm signal and its power spectrum} \label{sec:bayesianrecovery}

In this section we construct Gibbs sampling implementations capable of sampling from the full joint posterior distribution of the signal model, its power spectrum, and a set of foreground parameters. We begin by describing our model for visibility data and writing down the posterior distribution for the signal, and then demonstrate some of the key components of the Gibbs sampler using a hierarchy of Bayesian methods of increasing complexity, running from a simple maximum a posteriori (MAP) solution, to sampling from the signal distribution conditional on the signal covariance, to sampling from the full joint posterior distribution with a Gibbs sampler. We then detail three implementations that differ in their handling of the foreground component. Our implementation structures and notation follow those in  \cite{Eriksen}.

\subsection{Data model and posterior} 
We model visibility data $\m{V}$, which are complex valued, as 
\begin{equation}
    V_{mn}(\nu,t) = w(\nu, t)\bigg[ s_{mn}(\nu,t) +n_{mn}(\nu,t)\bigg],
\end{equation}
where the indices $m,n$ label the antennas used to form each visibility; $\nu$, $\tau$ and $t$ label frequency, delay (the Fourier conjugate to frequency), and observation time respectively; $\m{w}$ is a mask vector with values of 1 (unflagged) or 0 (flagged); $\m{n}$ is a Gaussian noise component with covariance $\m{N} \equiv \langle \m{n} \m{n}^\dag \rangle $; and $\m{s}$ is the total (signal + foreground) component in the data space. It is also useful to express the signal component in the discrete Fourier transform (DFT) basis as $\m{s} = \m{T \tilde s}$, where $\m{T}$ is a DFT matrix operator\footnote{$T_{mn} = e^{-2\pi i mn / N_{\textrm{freq}}}$. We use a Fourier convention such that $\m{T}^\dag\m{T} = \m{I}$, where the $\dag$ symbol denotes the Hermitian (conjugate) transpose.} and $\m{\tilde s}$ are coefficients of the total signal + foreground component in the Fourier basis. We use bold symbols to denote vector quantities, and upper-case letters to denote matrices. In general the total signal will be comprised of multiple components such as foregrounds, which we label $\m{f}$, the 21cm signal, labelled $\m{e}$, and instrumental systematics (which we neglect here). We assume that these components carry independent information, so that the total signal $\m{s}$ and the total signal covariance $\m{S}$ can be written
\begin{align}
    &\m{s} = \m{e} + \m{f} \\
    &\m{S} \equiv \langle \m{s} \m{s}^\dag \rangle = \langle \m{e} \m{e}^\dag \rangle + \langle \m{f} \m{f}^\dag \rangle = \m{E} + \m{F}. \label{eq:covE}
\end{align}
We would now like to find a way to estimate the joint posterior distribution of the model, $p(\m{s},\m{S}, \boldsymbol{\theta}| \m{d})$, conditioned on the measured data $\m{d}$, across all components of the signal and their covariance, and any other parameters $\boldsymbol{\theta}$ that may be considered in the analysis. The joint posterior contains not just estimates for $\m{s}$ and $\m{S}$, but also complete information about statistical uncertainties and correlations between parameters. In our case, it is a function of a large number of parameters however, including the values of the signal/foreground visibilities at each frequency channel, time, and baselines, and the elements of each covariance matrix. It would therefore be prohibitively expensive to explore the posterior directly due to the large number of dimensions. In the next sections, we outline three approaches of increasing complexity to handle the Bayesian estimation of the parameters in our model.

\subsection{Maximum a posteriori solution (Wiener filter)}
As a first step, consider the posterior distribution conditional on known covariance information. We consider the case of a single baseline and drop the antenna labelling indices $m,n$. Using Bayes' theorem, the signal's posterior distribution conditional on known covariances  $\m{S},\m{N}$ and measured data $\m{d}$ is
\begin{equation}
    p\big( \m{s}| \m{S},\m{N},\m{d}\big) \propto p\big(\m{d} |\m{s}, \m{S},\m{N}\big) p\big(\m{s} | \m{S}\big). \label{post}
\end{equation}
The second RHS distribution is a prior term for the signal $\m{s}$ given the data-space signal covariance, $\m{S}$, which will generally be independent of the data and the noise covariance. 
We assume our noise to be Gaussian distributed, which leads to the following conditional distribution:\footnote{Note that these are complex Gaussian distributions, which do not have a factor of $\frac{1}{2}$ in the exponent (or a square root of the determinant in the normalisation factor) when written in complex vector form \citep{Gallager2013}.\label{footnote_complex}}
\begin{align}
   p\big( \m{s} | \m{S},\m{N},\m{d}\big) \propto  e^{ - \left(\m{d} - \m{s} \right)^\dag \m{N}^{-1} \left(\m{ d} - \m{s} \right)}  e^{-\m{s}^\dag \m{S}^{-1}\m{s}}. \label{posterior}
\end{align}
Under the assumption of Gaussianity, the maxima of the posterior and log-posterior occur at the same location. To obtain the MAP estimate of the signal $\m{\hat s}$, the first derivative of the logarithm of $p\big(\m{s} | \m{S},\m{N},\m{d}\big)$ can be set to zero,
\begin{equation}
    \frac{\partial}{\partial \m{s}} \Bigg|_{\m{s}=\m{\hat s}} \bigg( \big(\m{d} - \m{s} \big)^\dag \m{N}^{-1} \big(\m{d} - \m{s} \big) + \m{s}^\dag \m{S}^{-1}  \m{s} \bigg) = 0,
\end{equation}
to obtain
\begin{equation}
    \m{d}^\dag \m{N}^{-1}  = \m{\hat s}^\dag \m{N}^{-1}  + \m{\hat s}^\dag \m{S}^{-1} .
\end{equation}
Using the fact that covariance matrices are Hermitian to remove complex conjugation, this equation can be rearranged to the `generalised Wiener filter' for the signal component $\m{s}$,
\begin{equation}
    \m{s}_{\textrm{wf}}  = \bigg[  \m{S}^{-1} + \m{N}^{-1}   \bigg]^{-1} \m{N}^{-1} \m{d}. \label{wf}
\end{equation}
In essence the Wiener filter answers the question: `given an assumption about the signal and noise covariance information, what is the most likely form of the realised signal in this data?'. Despite being the maximum a posteriori solution for the signal component of a data vector, the expectation of the solution $\langle\m{s}_{\textrm{wf}}\rangle \neq \m{s}$ in general \citep[e.g.][]{1992ApJ...398..169R}, though this bias can be partially ameliorated. The covariance of the Wiener filter estimate is also strictly smaller than the covariance of the parent conditional distribution. These biases are discussed in Appendix~\ref{sec:appA}. Defining the Wiener filter operation from Eq.~\ref{wf} as 
\begin{equation}
    \m{G} \equiv \bigg[ \m{S}^{-1} + \m{N}^{-1} \bigg]^{-1}\m{N}^{-1},
\end{equation}
and returning to the conditional Eq.~\ref{posterior}, the prior and likelihood terms can be combined by completing the square to form a single Gaussian
\begin{equation}
   p\big( \m{s} | \m{S},\m{N},\m{d}\big) \propto  \exp\bigg( - \big(\m{s} - \m{G}\m{d} \big)^\dag \big(\m{S}^{-1}+\m{N}^{-1}\big) \big(\m{s} - \m{G}\m{d} \big) \bigg) \label{posteriorcr},
\end{equation}
where the matrix identity $\m{A} + \m{B} = \m{A} \big[ \m{A}^{-1} + \m{B}^{-1} \big]\m{B}$ has been used, and data-only terms that are constant in the posterior have been absorbed as constants of proportionality. Here we see how the Wiener-filtered data $\m{s}_{\textrm{wf}} = \m{Gd}$ represents the mean of the distribution, and therefore that the Wiener filter minimises the expected residual $\langle|\m{s}  - \m{Gd}|\rangle$ between the true signal and the signal estimate.\footnote{The Wiener filter yields equivalent estimates to Gaussian Process Regression (GPR) for a given covariance matrix/kernel \citep{sarkka2013continuous}, hence the similarities between the equations here and the ones for GPR methods \citep[e.g.][]{2021MNRAS.501.1463K}.} We also see that the conditional distribution has covariance $\big[\m{S}^{-1} + \m{N}^{-1}\big]^{-1}$.

The data contains flagged regions which have been set to zero due to the presence of RFI. Since we do not have access to the information underneath the RFI mask, we set the noise variance in those regions to infinity. In implementation terms, this can be handled by using an amended form of the inverse noise covariance $\m{\tilde N}^{-1}$ 
\begin{equation}
    \m{\tilde N}^{-1} = \m{w} \m{w}^{\rm T} \circ \m{N}^{-1} ,
\end{equation}
where $\bf{w}$ is the mask vector and $\circ$ denotes elementwise multiplication. Substituting $\m{\tilde N}^{-1}$ for $\m{N}^{-1}$ in Eq.~\ref{wf} amounts to zeroing the contribution from the data inside the mask ($\m{N}^{-1}\m{d}$ term). The inverse signal covariance term does {\it not} typically go to zero in these regions; the prior `takes over' signal estimation in lieu of information from the likelihood function.  We should clarify that this is {\it not} simply a matter of filling in the masked regions with draws from the prior -- the Wiener filter inside the masked regions is constrained to match up with the solution in the unmasked regions, and close to the mask boundary both the prior and the data contribute. 

\begin{figure*}
    \centering
    \includegraphics[width=2.1\columnwidth]{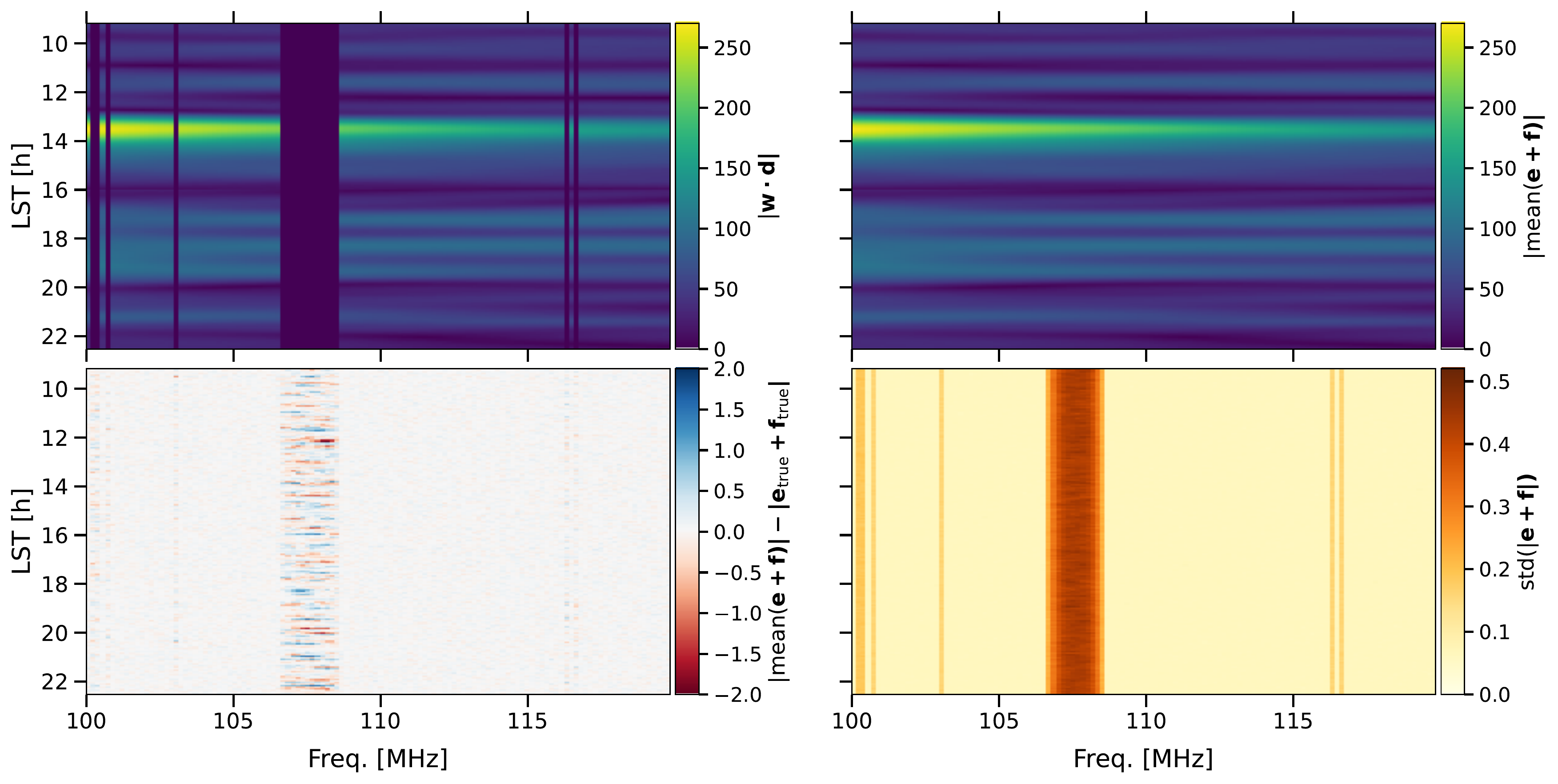}
    \caption{An example set of in-painted visibilities using Scheme 2, for a single 14.6m E-W baseline. 15\% of the band has been flagged, with a combination of a broad region and several randomly-flagged channels. The units of the visibilities are Jy.
    {\it Upper left:}  Amplitude (absolute value) of the `observed' visibilities, including flagged regions.
    {\it Upper right:} Amplitude of the mean of the EoR plus foreground realisations from the sampler (i.e. the average over all 800 samples).
    {\it Lower left:}  Difference between the mean of the EoR plus foreground realisations (upper right panel) and the {\it true} (input) EoR plus foreground model used in the simulation.
    {\it Lower right:} Standard deviation of the amplitude of the EoR plus foreground realisations from the sampler (calculated over all 800 samples).}
    \label{fig:waterfall}
\end{figure*}

\subsection{Gaussian constrained realisations from the conditional distribution of the EoR signal}

Having the MAP solution for the signal component in hand, the next step in Bayesian hierarchy is accessing the full conditional distribution $p(\m{s}|\m{S},\m{N},\m{d})$ and generating samples from it. To generate samples from the conditional, we make use of the Gaussian constrained realisation (GCR) equation, which we now describe. Since Eq.~\ref{posteriorcr} defines a multivariate Gaussian distribution, realisations drawn from the distribution (denoted $\m{s}_{\textrm{cr}}$) may be generated by adding random normal (Gaussian) realisations of the signal and noise fluctuation terms to the Wiener filter equation that are scaled correctly by their respective covariances \citep{Eriksen}, i.e.
\begin{equation}
    \m{s}_{\textrm{cr}}  = \bigg[ \m{S}^{-1} + \m{N}^{-1} \bigg]^{-1}  \bigg[\m{N}^{-1} \m{d} + \m{S}^{-1/2} \boldsymbol{\omega}_0 + \m{N}^{-1/2} \boldsymbol{\omega}_1  \bigg]. \label{cr}
\end{equation}
The new terms in the right bracket as compared with Eq.~\ref{wf} are independent realisations of zero mean, unit variance Gaussian random vectors $\boldsymbol{\omega}$ scaled by the noise and signal covariances.\footnote{Since the visibility data are complex-valued, so too must be the random vectors $\boldsymbol{\omega}$. In order to ensure that they have unit variance, we sum draws for unit variance real and complex parts, then divide this sum by $\sqrt{2}$.} With this form, it is straightforward to check that the covariance of a GCR solution is equal to the covariance of the Gaussian distribution in Eq.~\ref{posteriorcr}. Many realisations of the signal consistent with the given data vector $\m{d}$ and covariances $\m{S},\m{N}$ can be generated by solving Eq.~\ref{cr} repeatedly with different random realisations of the vectors $\boldsymbol{\omega}$, and these realisations trace the full conditional distribution.

Depending on the size of the data and the particular forms of the covariance matrices, solving Eqs.~\ref{wf} and \ref{cr} may become computationally demanding. A useful technique is to `precondition' the linear system by multiplying through by an easily-computable factor that makes the linear operator closer to the identity matrix (which would give a trivial solution to the linear system). In other words, for a linear system $\m{A} \m{x} = \m{b}$, preconditioned as $\m{P} \m{A} \m{x} = \m{P} \m{b}$, the best preconditioning matrix $\m{P}$ is one that gives $\m{P}\m{A} \approx \m{I}$, and where computing $\m{P}\m{A}$ is quite fast. As pointed out by \citet{Eriksen}, an effective preconditioner for the Wiener filter and GCR equations can be obtained by multiplying through by $\m{S}^{1/2}$, enabling much faster convergence in practice. With this preconditioning scheme, Eq.~\ref{cr} becomes
\begin{align}
   &\bigg[ \m{I} + \m{S}^{1/2} \m{\tilde N}^{-1} \m{S}^{1/2} \bigg] \m{y}_{\textrm{cr}}  = \m{S}^{1/2} \m{\tilde N}^{-1} \m{d}    +  \boldsymbol{\omega}_0 + \m{S}^{1/2}\m{\tilde N}^{-1/2} \boldsymbol{\omega}_1.\label{cry}   
\end{align}
The $\m{y}_{\textrm{cr}}$ solution vector can be obtained using a conjugate gradient solver, from which solution for the signal component can then be found as $\m{s} = \m{S}^{1/2} \m{y}_{\textrm{cr}}$.

Illustrative results from the GCR solver are shown in Fig.~\ref{fig:waterfall}, with a comparatively large flagged region close to the middle of the band, plus some randomly flagged channels, for a total flag fraction of 15\%. This example has been run as part of a full Gibbs scheme (Scheme 2; see Sect.~\ref{sec:scheme2} below), which runs for 800 iterations and includes sampling of the EoR and foreground covariance matrices/model parameters. The upper right panel of Fig.~\ref{fig:waterfall} shows an essentially seamless in-painting, with no visible discontinuity between the unflagged and in-painted regions. This continuity is due to the solution inside the flagged region being conditioned on the data outside, as well as the frequency structure of the EoR/foreground models and their covariances.

The lower left panel of Fig.~\ref{fig:waterfall} shows the difference between the absolute value of the mean (over 800 samples) of the GCR-sampled EoR plus foreground model with the true (input) EoR plus foreground model used in the simulation. The difference is small outside the flagged regions, as expected when fitting an accurate model to data with high SNR (for reference, the mean noise rms on the real and imaginary parts of the visibilities is 0.065, whereas the input EoR signal has an rms of 0.52 in the units of Fig.~\ref{fig:waterfall}). It is larger inside the flagged regions, but only shows structure that is non-smooth with frequency. This is also expected -- the data outside the flagged region constrain the possible behaviours of the EoR signal plus foregrounds inside the region, but do not fully specify them, particularly as one moves further from the edge of the region. The EoR signal in particular is allowed to be non-smooth in frequency, and so the solution inside the flagged region need not be strongly correlated with the solution outside. In any case, there is no clear bias towards either over- or under-estimating the combined EoR plus foreground signal inside the flagged region; as we will show later, the power spectrum of the EoR component in particular is recovered in an unbiased manner.

The lower right panel of Fig.~\ref{fig:waterfall} shows the variability in the EoR plus foreground realisations, plotted as the standard deviation of the absolute value of these quantities over 800 samples. It can be seen that the variability is lower for narrow flagged regions, as only relatively smaller deviations are allowed in order for the solution to remain consistent with the surrounding data. The variability is larger inside the broad flagged region, but shows a smooth transition from the edge to the centre, again due to the solution being constrained more strongly by neighbouring data near the edge of the region. Note that the variability of the realisations is also non-zero outside the flagged regions. This is driven mostly by the noise level; neither the EoR nor foreground solutions are completely certain in the unflagged regions, even though the SNR is reasonably large.

\subsection{Realisations from the joint posterior (Gibbs)}

Gibbs sampling is a method for recovering the joint posterior distribution, in our case $p(\m{s},\m{S}|\m{d})$, via MCMC sampling \citep{geman1984stochastic}. Under the condition that a given joint probability density is strictly positive across the span of each variable (i.e. that no point in the joint space has zero probability density), then that joint density is specified uniquely by the full set of conditional distributions for all the parameters. Gibbs sampling uses this fact by sampling from each conditional distribution in turn, in the process updating conditioned-on variables with the sample obtained for them at the previous iteration. Since the joint posterior we wish to evaluate is a function of two (vector) quantities, the signal $\m{s}$ and the signal covariance $\m{S}$, the joint posterior can be evaluated by sampling (indicated by $\leftarrow$) from each of the two conditional distributions iteratively,
\begin{align}
    &\m{s}_{i+1} \leftarrow  p(\m{s}_{i}|\m{S}_i,\m{N},\m{d}) \label{eq:conditional_s}\\
    &\m{S}_{i+1} \leftarrow  p(\m{S}_{i} | \m{s}_{i+1}). \label{eq:conditional_S}
\end{align}
In the above representation, the top line is sampled first, followed by the second line, with the index $i$ running over iterations. The distribution for the signal covariance $\m{S}$ is not conditioned on the noise covariance $\m{N}$ and data $\m{d}$ since all of the relevant information is contained in the current realisation of $\m{s}$. Eq.~\ref{eq:conditional_s} has the form of a multivariate Gaussian, and so sampling is achieved by making use of the GCR equation, Eq.~\ref{cr}. The conditional distribution of the covariance, Eq.~\ref{eq:conditional_S}, on the other hand, has the form of a complex inverse Wishart distribution,
\begin{equation}
    p(\m{S}|\m{s}) = \frac{p(\m{s}|\m{S})p(\m{S})}{p(\m{s})} \propto \frac{1}{{\textrm{det}(\m{S})}} \exp\bigg(- \m{s}^\dag \m{S}^{-1} \m{s} \bigg), \label{invw}
\end{equation}
where det$(...)$ signifies the matrix determinant (recall that there is no square root of the determinant when written in complex vector notation; see footnote \ref{footnote_complex}). Since we solve for full realisations of the signal component at the GCR step (i.e. that have values both inside and outside regions masked due to RFI), the need for an in-painting process that explicitly estimates the missing data is eliminated.

In the next three subsections, we detail a series of Gibbs sampling implementations that treat the foreground component in different ways.

\subsection{Scheme 1 (combined signal + foreground sampler)}

The first scheme is a straightforward implementation sampling the total signal from a complex Gaussian distribution ($\mathcal{CN}$) using the GCR equation (i.e. sampling $\m{s} = \m{e} + \m{f}$), followed by sampling the total covariance matrix $\m{S}$ from the complex inverse Wishart ($\mathcal{CW}^{-1}$) distribution,
\begin{align}
    &\m{s}_{i+1} \leftarrow \mathcal{CN}\bigg(\m{Gd},\big[\m{S}_i^{-1} + \m{N}^{-1}\big]\bigg)\label{cn}\\
    &\m{S}_{i+1} \leftarrow \mathcal{CW}^{-1}\bigg( \Sigma_{i+1}, \nu_f, N_{\textrm{freq}}\bigg),\label{cinvw}
\end{align}
where the sample covariance matrix is
\begin{equation}
\Sigma_{i+1} = \frac{1}{N_{\rm times} - 1} \sum_t^{N_{\rm times}} \m{s}_{t,i+1}^{\,} \m{s}_{t,i+1}^{\dag},
\end{equation}
the parameter $\nu_f = N_{\textrm{freq}}(N_{\textrm{freq}}+1)/2$ denotes the number of degrees of freedom in the sample covariance, and $N_{\rm times}$ is the number of LSTs in the visibility data. We discuss obtaining samples from a complex inverse Wishart distribution in Appendix~\ref{sec:appB}. In each iteration of the Gibbs sampler the GCR equation (Eq.~\ref{cn} above) is solved using visibility data across all LSTs separately, using as a prior the covariance matrix arrived at during the last iteration. The mean covariance used for sampling from the distribution in Eq.~\ref{cinvw} is the sum over outer products of the obtained GCR solutions from each LST at the previous step. This sampling scheme will arrive at realisations of both the total signal and total data covariance in frequency space.

\subsection{Scheme 2 (joint sampler with foreground templates)} \label{sec:scheme2}

In the second implementation, we make two simplifying assumptions: i) the delay spectrum of the EoR signal component is diagonal, i.e. all of the statistical information about $\m{s}$ is contained by its power spectrum\footnote{In making this assumption we automatically lose sensitivity to any non-Gaussian information in the measured EoR field. In practice, that means that such an assumption is fairly limiting since the EoR field is expected to be significantly non-Gaussian. Testing the extent to which this assumption limits us will require more sophisticated simulations than we employ in this work.}, and ii) that the foreground covariance matrix $\m{F}$ is known, and hence fixed (not sampled). This implementation follows a Gibbs sampling scheme for the CMB signal map and power spectrum \citep{Eriksen}. Foreground covariance matrices can be estimated directly from simulations based on past observations (see Sect.~\ref{sec:sims}), and as such the foreground covariance is known to a higher degree of accuracy than the covariance of the EoR field. Fixing this quantity to the values arrived at via simulations significantly reduces the volume of parameter space to be covered by the sampler. Secondly, the cosmological signal $\m{e}$ having a diagonal covariance in delay space (i.e. with variance given by delay spectrum in each delay bin) reduces the complex inverse Wishart distribution of Eq.~\ref{cinvw} to a product of inverse Gamma distributions, one for each delay spectrum bandpower.

\begin{figure}
    \centering
    \includegraphics[width=1.0\columnwidth]{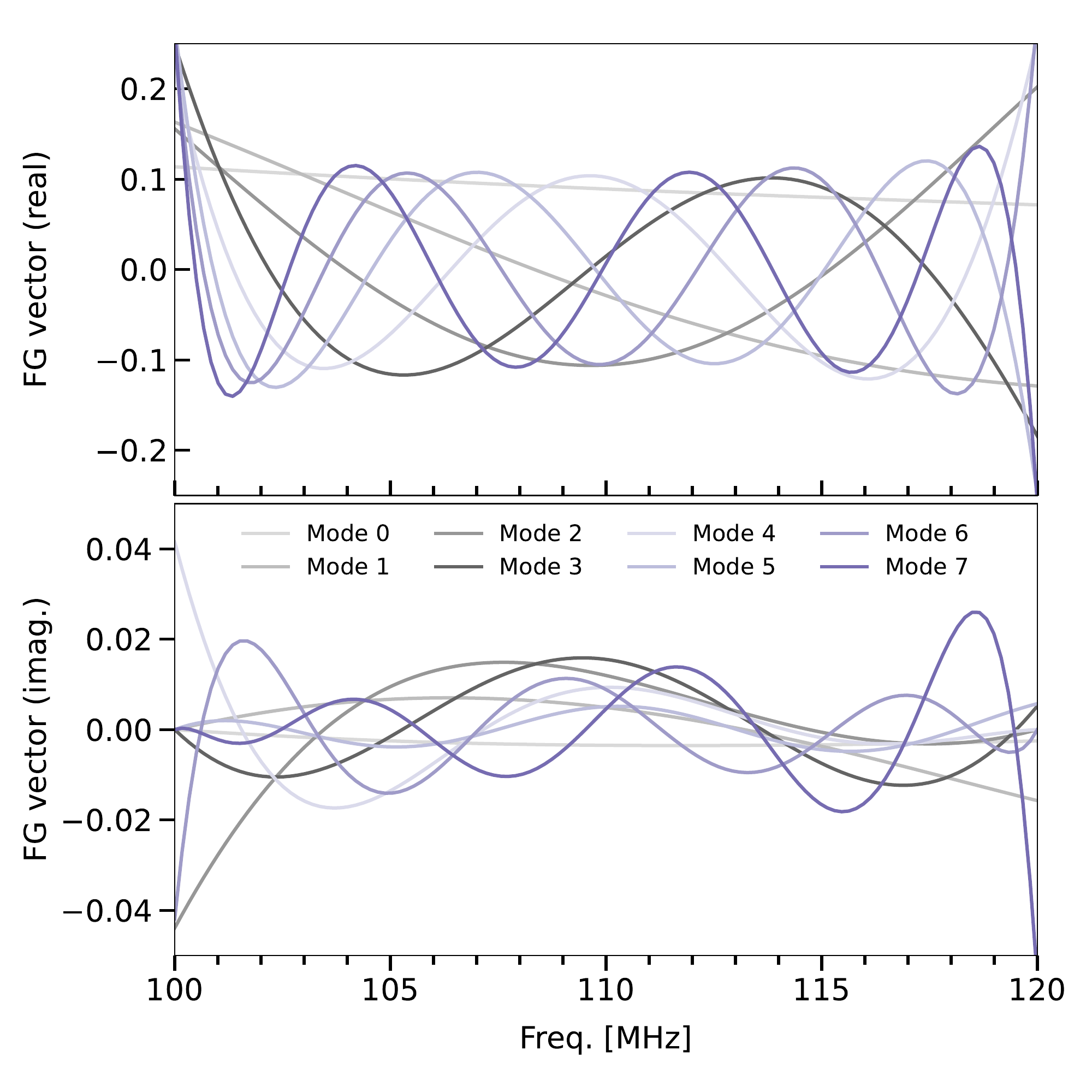}
    \caption{First 8 foreground principal components (covariance eigenmodes) derived from the frequency-frequency covariance matrix measured from simulations of a 14.6m E-W baseline. The upper and lower panels show the real and imaginary parts respectively.}
    \label{fig:emodes}
\end{figure}

As with the implementation in \citet{Eriksen}, we make use of a set of foreground templates $\m{g}_j(\nu)$ with respective amplitudes $\m{a}_{\textrm{fg}}$. For our templates, we use the first $N_{\textrm{pc}}$ principal components of our simulation-derived foreground covariance matrix on each baseline such that the (non-square) template matrix has dimension $N_{\textrm{freq}} \times N_{\textrm{pc}}$. An example set of principal component modes is shown in Fig.~\ref{fig:emodes}. The data model in this scheme is given by
\begin{equation}
    \m{d} = \m{e} + \m{g}_j \cdot \m{a}_{\textrm{fg}} + \m{n},
\end{equation}
where $\m{e}$ denotes the cosmological signal in frequency space.
We follow \citet{Eriksen} in defining a vector $\m{x}$ where the first $N_{\textrm{freq}}$ entries are $\m{e}$ and the second $N_{\textrm{pc}}$ entries are $\m{a}_{\textrm{fg}}$, and a corresponding response vector $\m{u} = (\m{1}, \m{g}_j)^{\rm T}$ so that the signal and foreground amplitude conditional distribution may be written
\begin{align}
    &p(\m{e} , \m{a}_{\textrm{fg}} | \m{E}, \m{g}_j, \m{N}, \m{d}) \propto p(\m{d} | \m{e}, \m{a}_{\textrm{fg}}, \m{g}_j, \m{N}) p(\m{e}|\m{E}) \nonumber\\
    &\propto \exp\bigg( - \big(\m{d} - \m{x}\cdot\m{u} \big)^\dag \m{N}^{-1} \big(\m{ d} - \m{x}\cdot\m{u} \big) \bigg)  \exp\bigg(- \m{s}^\dag \m{S}^{-1}\m{s}  \bigg)\nonumber\\
    & \propto \exp\bigg( - \big(\m{x} - \m{\hat x} \big)^\dag \m{A}^{-1} \big(\m{x} - \m{\hat x}\big)\bigg),
\end{align}
where the form of the vector $\m{\hat x}$ and matrix $\m{A}$ follow from a completed-square representation of the line above. 
The symbolic linear system to be solved, $\m{A}\m{x}=\m{b}$, comparable to Eq.~\ref{cr}, takes the explicit form
\begin{align}       
\begin{bmatrix}
\m{E}^{-1} + \m{N}^{-1}& \m{N}^{-1} \m{g}_j\\
\m{g}_j^\dag \m{N}^{-1}  & \m{g}_j^\dag  \m{N}^{-1} \m{g}_j 
\end{bmatrix}
\begin{bmatrix}
\m{e}\\
\m{a}_{\textrm{fg}}
\end{bmatrix} ~~~~~~~~~~~~~~~~~~~
\nonumber \\
~~~~~~~~~~~~~~ = 
\begin{bmatrix}
\m{N}^{-1} \m{d} + \m{E}^{-1/2} \boldsymbol{\omega}_0 + \m{N}^{-1/2} \boldsymbol{\omega}_1 \\
\m{g}_j^\dag \m{N}^{-1} \m{d} + \m{g}_j^\dag \m{N}^{-1/2} \boldsymbol{\omega}_1 
\end{bmatrix}, \label{scheme2} 
\end{align}
where the foreground model can be recovered as $\m{f} = \m{g}_j\cdot\m{a}_{\textrm{fg}}$, and $\m{E}$ is the EoR signal covariance in frequency space. The sampling is performed for each LST independently and in parallel, and so we obtain a set of $N_t$ solution vectors $\m{e}$ and $\m{a}_{\rm fg}$ at each iteration.

The bandpowers of the EoR signal delay spectrum $\m{p} = (P(\tau_0), \ldots, P(\tau_N))^{\rm T}$ for delay bins $\tau_0 \ldots \tau_N$ are related to the frequency-space EoR covariance by $\m{E} = \m{T}\m{\tilde{E}}\m{T}^\dag$, where the delay-space covariance $\m{\tilde{E}}$ is zero everywhere except on the diagonals, where ${\rm diag}(\m{\tilde{E}}) = \m{p}$. Each bandpower is sampled independently from an inverse Gamma distribution with a scale parameter calculated from the variance (over LST) of the current realisation of the corresponding EoR signal delay mode, i.e. the variance for delay mode $\tau$ at iteration $i+1$ is 
\begin{equation}
\sigma^2_{\tau,i+1} = \frac{1}{N_{\rm times}-1}\sum_{t}^{N_{\rm times}} \tilde{e}_{\tau,t,i+1}^* \tilde{e}_{\tau,t,i+1}, \label{eq:sigmatau}
\end{equation}
where $\m{\tilde{e}}_t = \m{T}^\dag \m{e}_t = (\tilde{e}_{\tau_0, t}, \ldots, \tilde{e}_{\tau_N, t})^{\rm T}$.
This Gibbs scheme therefore has two steps,
\begin{align}
    &\m{e}_{i+1}, \m{a}_{\textrm{fg},i+1} \leftarrow p(\m{e}_i , \m{a}_{\textrm{fg},i} | \m{E}_i, \m{g}_j, \m{N}, \m{d});\nonumber \\
    & P(\tau)_{i+1} \leftarrow \textrm{Inv-Gamma}\bigg( \sigma^2_{\tau,i+1}, \alpha \bigg),
    \label{s2}
\end{align}
where the joint amplitude sampling is carried out for each LST separately, the bandpower sampling is performed separately for each delay mode, and $\alpha = N_{\textrm{visibilities}} -1$. In words, the above scheme jointly samples the EoR signal vector (in frequency space) and amplitudes of the foreground templates $\m{g}_j$, for each LST independently, by solving the linear system of Eq.~\ref{scheme2}. It then calculates the sample variance of each EoR signal delay mode over all available LSTs, and uses this as a scale parameter to draw a sample of each delay spectrum bandpower from an inverse-Gamma distribution, independently for each delay mode. It then finally performs an inverse Fourier transform of a delay-space covariance matrix constructed by putting the delay spectrum bandpower samples along the diagonal (and zeros elsewhere), resulting in the $i+1^{\textrm{th}}$ sample of the frequency-space signal covariance, $\m{E}_{i+1}$. Due to our enforcement of the property that $\m{E}$ be diagonal in delay space, it is necessarily a circulant matrix in frequency space \citep{messerschmitt2006}.

The delay spectra of the foregrounds and the EoR signal are degenerate where they overlap at low delay. This is due to the foreground modes corresponding to smooth functions in frequency that are representable with only a handful of low-wavenumber Fourier modes. We implement a symmetric prior on a few low delay bins of $\m{p}$ to control the degeneracy. Specifically, we set the result of the delay spectrum (inverse-Gamma) sample $P(\tau)_{i+1}$ to be equal to the true EoR delay spectrum $P_{\rm true}(\tau)$ inside delays $-100~ {\rm ns} < \tau < 100~\rm{ns}$. This prior prevents time-consuming exploration of the degeneracy which would slow convergence of the chains.

It is important to note that without this highly specific prior, the sampler is unable to disambiguate between the smoothest foreground modes and the EoR signal, meaning that the lowest-delay EoR signal modes will not be recovered correctly. In the case of real data, a different prior, for example a continuity prior, is likely to be more suitable as the true EoR delay spectrum is unknown. One could also use a set of physically-motivated models to constrain possible behaviors of the EoR signal at low delay. We do not explore these possibilities further here.

\subsection{Scheme 3 (joint sampler with signal and foreground constrained realisations)}

This scheme jointly samples from the EoR signal and foreground components $\m{e}$ and $\m{f}$ at the GCR step using the full foreground covariance matrix $\m{F}$, rather than its leading principal components as in Scheme 2. The assumption is maintained that the foreground covariance does not vary.\footnote{In principle, this scheme could be extended to also sample the foreground covariance matrix $\m{F}$, e.g. as another inverse Wishart Gibbs step, but we leave this to future work.} Again defining a block vector $\m{x}$ where now the first $N_{\textrm{freq}}$ entries are $\m{e}$ and the second $N_{\textrm{freq}}$ entries are $\m{f}$, and a corresponding response vector $\m{u} = (\m{1}, \m{1})^{\rm T}$ such that the data model is $\m{d} = \m{x} \cdot \m{u} + \m{n}$. From the joint conditional distribution of the signal and foreground vectors, we obtain
\begin{align}
    &p(\m{e} , \m{f} | \m{E}, \m{F}, \m{N}, \m{d}) \propto p(\m{d} | \m{e}, \m{f}, \m{N}) p(\m{e}|\m{E}) p(\m{f}|\m{F})\nonumber\\
    &\propto \exp\bigg( - \big(\m{d} - \m{x}\cdot\m{u} \big)^\dag \m{N}^{-1} \big(\m{ d} - \m{x}\cdot\m{u} \big) \bigg)\nonumber\\
    & ~~~\times \exp\bigg(-  \m{e}^\dag \m{E}^{-1}\m{e}\bigg) \exp\bigg(-  (\m{f-\bar{f}})^\dag \m{F}^{-1}(\m{f-\bar{f}})  \bigg), \label{scheme3}
\end{align}
with $\m{\bar{f}}$ being the mean vector of the foregrounds on this baseline. Completing the square with the vector $\m{x}$, the corresponding linear scheme to solve to sample from the conditional distribution is 
\begin{align}       
\begin{pmatrix}
\m{E}^{-1} + \m{N}^{-1}& \m{N}^{-1} \\
 \m{N}^{-1}  & \m{F}^{-1} + \m{N}^{-1}  
\end{pmatrix}
\begin{pmatrix}
~\m{e}~ \\
~\m{f}~
\end{pmatrix}
 = \m{b},
\end{align}
with $\m{b}$ given by
\begin{align} \m{b} = 
\begin{pmatrix} 
\m{N}^{-1} \m{d} + \m{E}^{-1/2} \boldsymbol{\omega}_0 + \m{N}^{-1/2} \boldsymbol{\omega}_1 \\
\m{N}^{-1} \m{d} + \m{F}^{-1}\m{\bar{f}} + \m{F}^{-1/2} \boldsymbol{\omega}_2 + \m{N}^{-1/2} \boldsymbol{\omega}_1
\end{pmatrix}.
\end{align}
This implementation samples from the foreground and signal components jointly. Since the foreground covariance on the baseline is taken to be known, it is not sampled from. The (baseline specific) foreground covariance matrix $\m{F}$ is a dense matrix in frequency space that needs to be inverted in order to use this implementation. This Gibbs sampling scheme is very similar to Scheme 2:
\begin{align}
    &\m{e}_{i+1}, \m{f}_{i+1} \leftarrow p(\m{e}_i , \m{f}_i | \m{E}_i, \m{F}, \m{N}, \m{d});\nonumber \\
    & P(\tau)_{i+1} \leftarrow \textrm{Inv-Gamma}\left( \sigma^2_{\tau,i+1}, \alpha \right),
\end{align}
with foreground vectors now being sampled jointly with the EoR signal realisations. As with Scheme 2, the signal realisations are used to form the basis for inverse-Gamma samples of the delay spectrum, which again is used to define a diagonal covariance matrix in delay space. We also implement the same prior on the central delay bins of $\m{p}$ as for Scheme 2.

\section{Simulations} \label{sec:sims}

\begin{figure*}
    {\centering
	\includegraphics[width=2.\columnwidth]{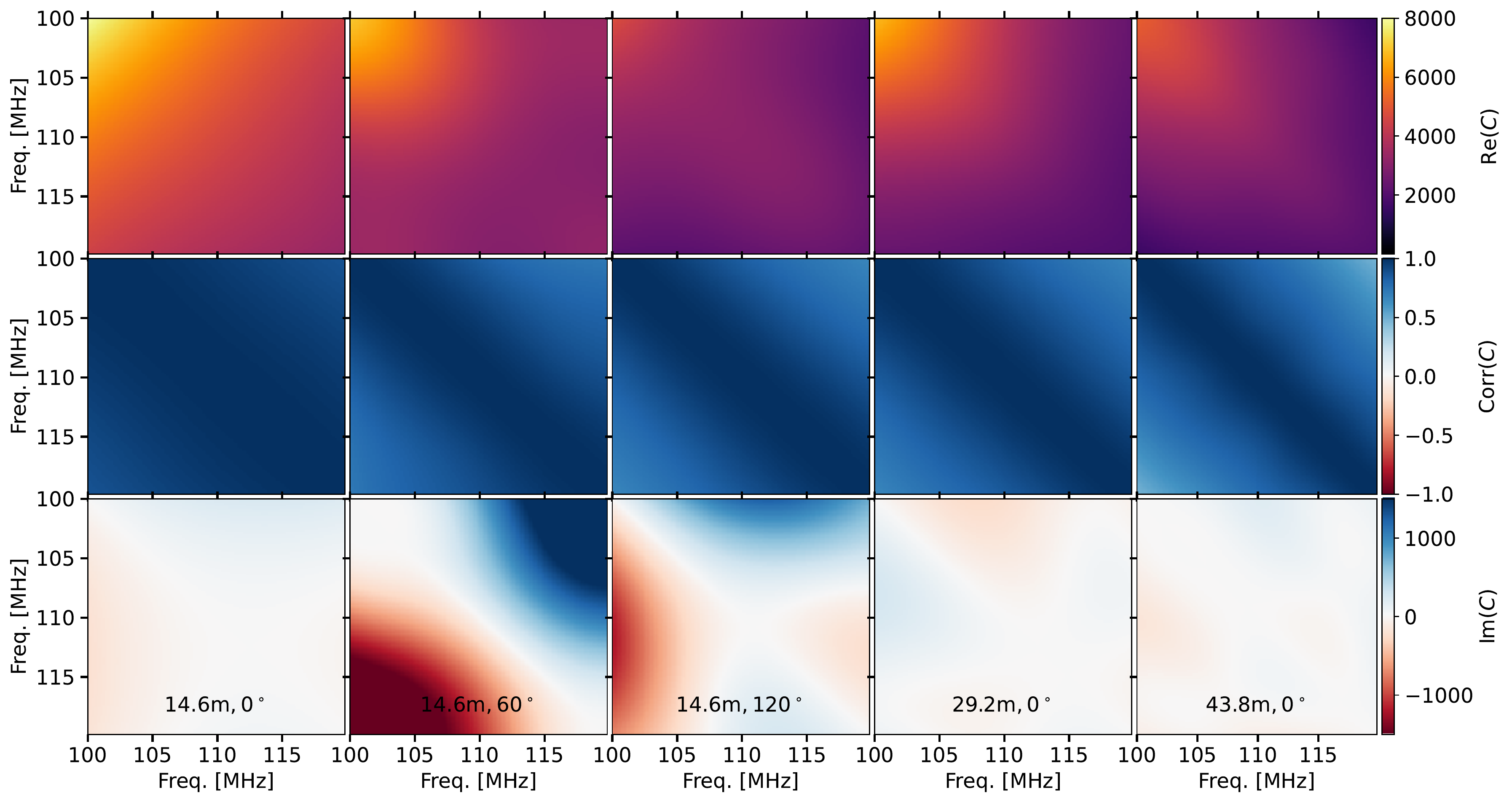}
	}
	\caption{Point source foreground covariance matrices estimated from averaging over 1200 simulation time samples at a cadence of 40s in the 100--120~MHz band. Real and imaginary parts (in Jy$^2$) are shown in the top and bottom rows respectively, while the middle row shows the correlation matrix of the real part, $\rho_{ij} = C_{ij} / \sqrt{C_{ii} C_{jj}}$. The first three columns are for baselines with length 14.6m, orientated at approximately $0^\circ$, $60^\circ$, and $120^\circ$ degrees from the E-W direction respectively (the shortest redundant baselines). A clear directional dependence is found in the structure and amplitude of the covariance matrix. The last two panels show covariances from baselines of length 29.2m and 43.8m (both E-W aligned). The real part of each of the covariances shows strong correlations, with an imaginary part that has a smaller variance in general.}
	\label{fgcovs}
\end{figure*}

In this section we describe the steps taken to build model covariance matrices for the foregrounds, and the covariance used as our EoR signal model. We use the simulation methodology described in \citet{2021MNRAS.506.2066C} in what follows. The simulations include separate sets of simulated visibilities for point sources, diffuse emission (which we have neglected here), and a simple model of the EoR. They cover a bandwidth of 100-120 MHz in 120 channels, and contain 13.4 hours of LST at 40 seconds integration time per sample, in the LST range 9.2 -- 22.5 h. Only the pseudo-Stokes I polarisation channel is simulated. The {\tt hera\_sim}\footnote{\url{https://github.com/HERA-Team/hera_sim/}} package is used to perform the visibility simulations themselves, with an analytic approximation to the HERA beam \citep{2021MNRAS.500.1232F, 2021MNRAS.506.2066C} used as the primary beam model, which is furthermore assumed to be identical between receivers.

To simulate point sources we choose a sky model based on the GLEAM catalogue \citep{hurley17}. We also include a few bright sources and {\it Fornax A} which are not present in the catalogue (see Table 2 of \citealt{hurley17}). There are some blank regions in the GLEAM catalogue (e.g., north of $+30^{\circ}$ declination, Galactic latitudes within $10^{\circ}$ of the Galactic plane, and a handful of localised areas such as the Magellanic Clouds) with no sources. We fill those gaps with sources taken from other parts of the sky, as described in \citet{2021MNRAS.506.2066C}. 

Covariance matrices for simulated point source foregrounds are estimated per-baseline by averaging in the LST (time) direction (ignoring any non-stationarity of the statistics of the foregrounds). For a set of mean-subtracted visibility data $\m{V}$ of dimension $N_{\textrm{freq}} \times N_{\textrm{times}} $, the frequency-frequency covariance matrix can be estimated as
\begin{equation}
    \m{F} = \frac{1}{N_{\textrm{times}}-1} \m{V} \m{V}^{\dag}. \label{covcalc}
\end{equation}
In Fig.~\ref{fgcovs} we show examples of point source foreground covariance matrices formed from different baseline lengths and orientations in the simulated array. The real parts (upper row) show strong correlations, and the imaginary parts (lower row) have structures that change with baseline length and orientation. In the first three columns we show the covariance matrix of point source foregrounds on baseline vectors of length 14.6m that form an equilateral triangle. Though the real part takes on a similar structure for each baseline type triangle, there are notable differences. The first panel shown (14.6m, $0^\circ$) has the largest magnitude (denoting higher variance) and has strong correlations over larger frequency separations than the second panel (14.6m, $60^\circ$). The third panel (14.6m, $120^\circ$) shows a magnitude of the real part that is approaching a factor of two smaller than the first panel. Differences in the imaginary part are more striking, with very different correlation structures visible in each.

We also show covariance matrices for longer baselines in the fourth and fifth panels, of length 29.2m and 43.8m respectively, both orientated E-W (0$^\circ$). These baselines have covariances with shorter correlation lengths than the equivalent 14.6m baseline, which is expected as longer baselines are more chromatic. The longer baselines also show reduced magnitude of the real part compared to the shorter baselines.

Various point source covariance calculations and fitting functions exist \citep[e.g.][]{santos05, murray2017, 2020MNRAS.495.2813G}, and this study of the per-baseline covariances seems to indicate that a one-size-fits-all approach is likely to fall short. The simulations of \cite{2021MNRAS.506.2066C} include very bright sources such as Cen~A, which can significantly increase the total observed power (and therefore the variance) as they transit for example. This implies that deviations from the assumptions we have made in calculating the covariance -- stationarity, and statistical homogeneity/isotropy -- can be significant. The orientation of the baselines (and therefore the fringes) on the sky as sources rotate through them at different rates (i.e. at different `fringe rates') will also contribute to the structure of the correlations in the real and imaginary parts. We leave attempts at systematization to future work, and for the results in later sections take the foreground prior covariance $\m{F}$ to be the covariance matrix evaluated using Eq.~\ref{covcalc} for the baseline that we make use of to generate mock data, an E-W oriented 14.6m baseline.

For our EoR signal model we use a Gaussian plus a constant offset in delay space. In frequency space, this is \begin{align}
    \m{E}_g(\nu,\nu') &=  A_s \exp \bigg( -\frac{1}{2}\frac{(\nu-\nu')^2}{\omega_s^2}\bigg) + r_s\delta_{\nu\nu'} ,
    \label{signalmodel}
\end{align}
where $\delta_{\nu\nu'}$ is a Kronecker delta function, and the parameters take constant values $A_s = 0.25$, $\omega_s = 0.5$, and $r_s = 0.025$. Realisations from this signal covariance model produce Gaussian fluctuations with a correlation length and amplitude parameterised by $\omega_s$ and $A_s$ respectively. $r_s$ controls the size of a diagonal (in frequency) component that also ensures that $\m{E}_g$ is positive-definite. This signal model produces a power spectrum that is a Gaussian function around delay zero, plus an offset that is constant in delay. 

\section{Results} \label{sec:results}

In this section we show results from each sampler over a substantial number of iterations (800) and under different simulated data scenarios, and compare the delay spectrum of the recovered EoR component with the true, input EoR delay spectrum. The primary statistic we consider is the distribution of recovered delay spectra across the set of iterations, which approximates the marginal posterior distribution of the EoR delay spectrum.

First, we show results from runs of each Gibbs sampling scheme under a fiducial testing setup with 5$\%$ random flagging applied in the same way to each LST, a SNR of 5 at high delay for each visibility, and the Gibbs sampler chain initialised at the true signal and signal covariance values to avoid a lengthy burn-in period. We then consider three different scenarios: how changes to the flagging fraction affect delay spectrum recovery; how the signal power spectrum recovery is affected by changes to the SNR; and how an incorrectly estimated foreground covariance matrix affects delay spectrum recovery (in this case, one formed from visibility simulations that have no faint sources below 15~Jy).

\subsection{Simulation realisations} \label{sec:realsims}

We generate our simulated data using the following method. A point source foreground simulation from a 14.6m E-W baseline covering 1200 LSTs (40 seconds spacing) is taken as a base (see Sect.~\ref{sec:sims}). This simulation is held fixed throughout, i.e. we do not use any other realisation of the foreground visibilities than that in the simulated data. On top of it, we add an independent complex Gaussian white noise draw to each time and frequency channel, plus a simple EoR signal component that we generate using 1200 independent complex Gaussian random draws from the signal frequency-frequency covariance matrix (Eq.~\ref{signalmodel}). The shape of the EoR signal power spectrum is not chosen to represent any particular physical model; instead, we use a Gaussian shape with the peak at $\tau = 0$~ns and a width of around 1000~ns, as it provides a simple but non-trivial shape to recover, and allows low-, intermediate-, and high-SNR regimes to be studied in the same power spectrum.

Since each time sample of the EoR signal is drawn separately, the 1200 LSTs each contain an independent realisation drawn from the underlying signal power spectrum. This is akin to each time sample being taken approximately one primary beam crossing time apart, such that a different patch of the sky has rotated into the mainlobe of the beam in its entirety. For HERA, this timescale is of order $\sim 1$ h depending on the observing frequency, and so the choice of independent samples of the EoR signal is clearly an idealisation. In reality, observations 40 seconds apart would see a very similar sky within the primary beam mainlobe, and so the signal realisations would be strongly correlated. Recent HERA analyses \citep[e.g.][]{2022ApJ...925..221A} have performed coherent averaging of visibilities over timescales of a few minutes (with fringe stopping to reduce decoherence), which effectively increases the separation between each (post-averaging) time sample. Coherent averaging over baselines may also be performed to increase SNR. We do not model either of these forms of averaging here, however.

At each iteration the samplers return 1200 EoR signal GCR solutions, one for each LST. We estimate the `empirical' delay spectrum of the samples at each LST separately, by multiplying the Fourier transform of the mean-subtracted, tapered GCR solution by its complex conjugate and then averaging across all LSTs in the iteration to obtain
\begin{equation}
\m{P}_{\rm empirical} = \left ( \hat{\sigma}_{\tau_0, i}^2, \ldots, {\hat{\sigma}}_{\tau_N, i}^2 \right )^{\rm T} \label{eq:Pempirical}
\end{equation}
for iteration $i$. We have used $\hat{\sigma}^2_{\tau,i}$ to denote the same quantity as in Eq.~\ref{eq:sigmatau}, but with the delay mode $\tilde{e}_\tau$ replaced by an equivalent quantity that was tapered and mean-subtracted before the Fourier transform.
Note that $\m{P}_{\rm empirical}$ is different from the delay spectrum estimate $P(\tau)$ that is obtained via sampling by the second step of the Gibbs scheme. In particular, $P(\tau)$ is subject to (inverse-Gamma) sample variance while $\m{P}_{\rm empirical}$ is not, and the calculation of $\m{P}_{\rm empirical}$ includes a tapering operation while $P(\tau$) does not. We run each sampler for 800 iterations, which is sufficient to achieve good convergence based on visual inspection of the traces of the chains.

\begin{figure}
    {\centering
	\includegraphics[width=\columnwidth]{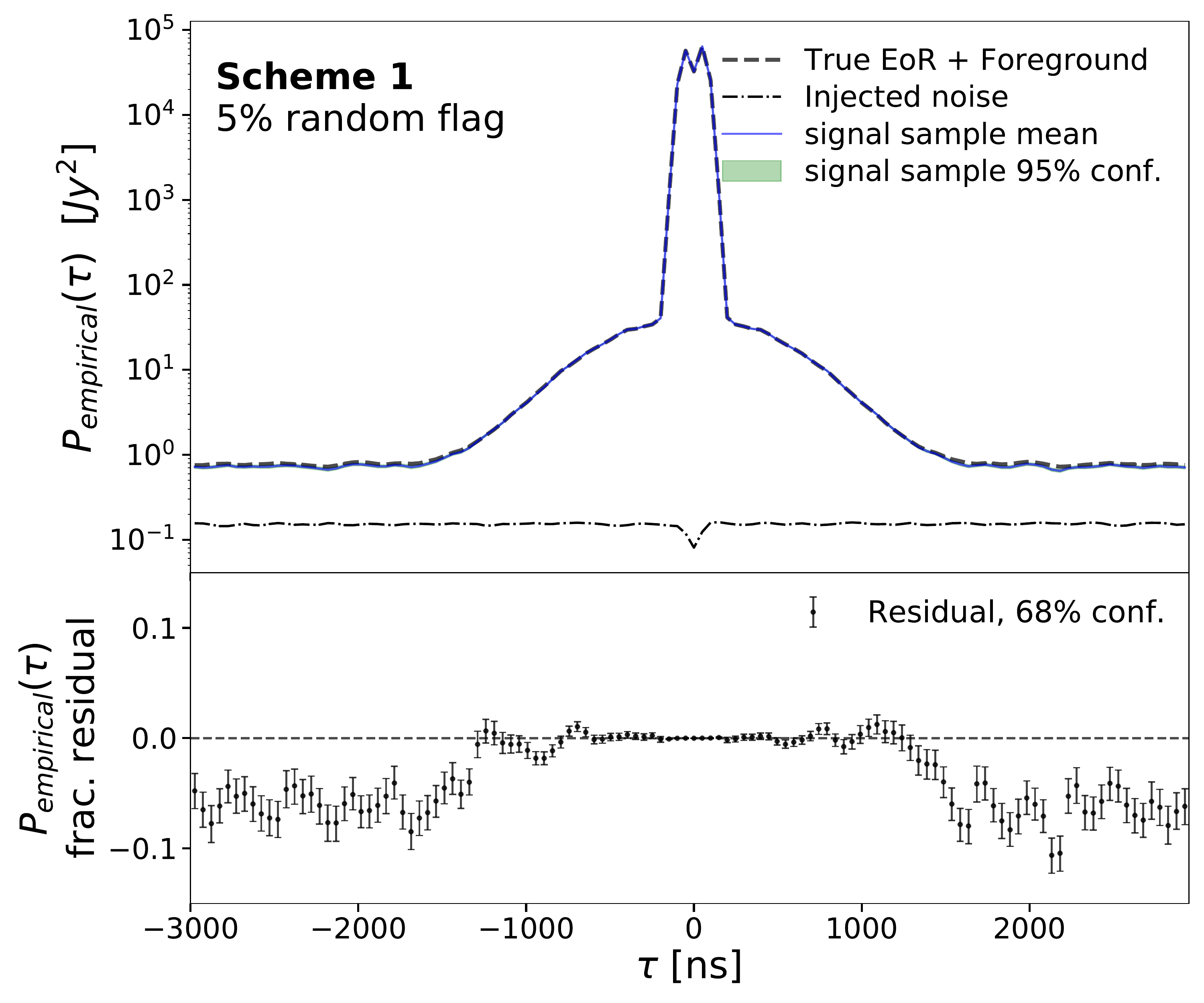}
	}
	\caption{Signal recovery using Scheme 1 (total signal sampler) after 800 iterations. {\it Top panel:} True input power spectrum (sum of EoR and foreground components, dashed line) compared with the mean of the power spectrum of the total sky signal estimated from samples of the signal from the GCR step (blue line).
	Power spectra are calculated after mean-subtraction and tapering with a Blackman-Harris window, which produces a small decrease at $\tau = 0$~ns. The dot-dashed line shows the noise power for each visibility (i.e. a single time sample; SNR of 5 at high $\tau$); 1200 time samples are combined to measure the total power spectrum. The 95\% confidence interval (green shaded region) is too small to see. {\it Bottom panel:} Fractional residual between the true power spectrum and the recovered power spectrum from the upper panel. The errorbars show the 68$\%$ confidence region. Outside $|\tau| \gtrsim 1000$~ns, where the SNR is lower, this sampling scheme recovers a distribution that is biased low by around 5$\%$. This bias does not appear when the SNR at high delay is increased.}
	\label{0-5-ci}
\end{figure}

\begin{figure*}
    {\centering
	\includegraphics[width=\columnwidth]{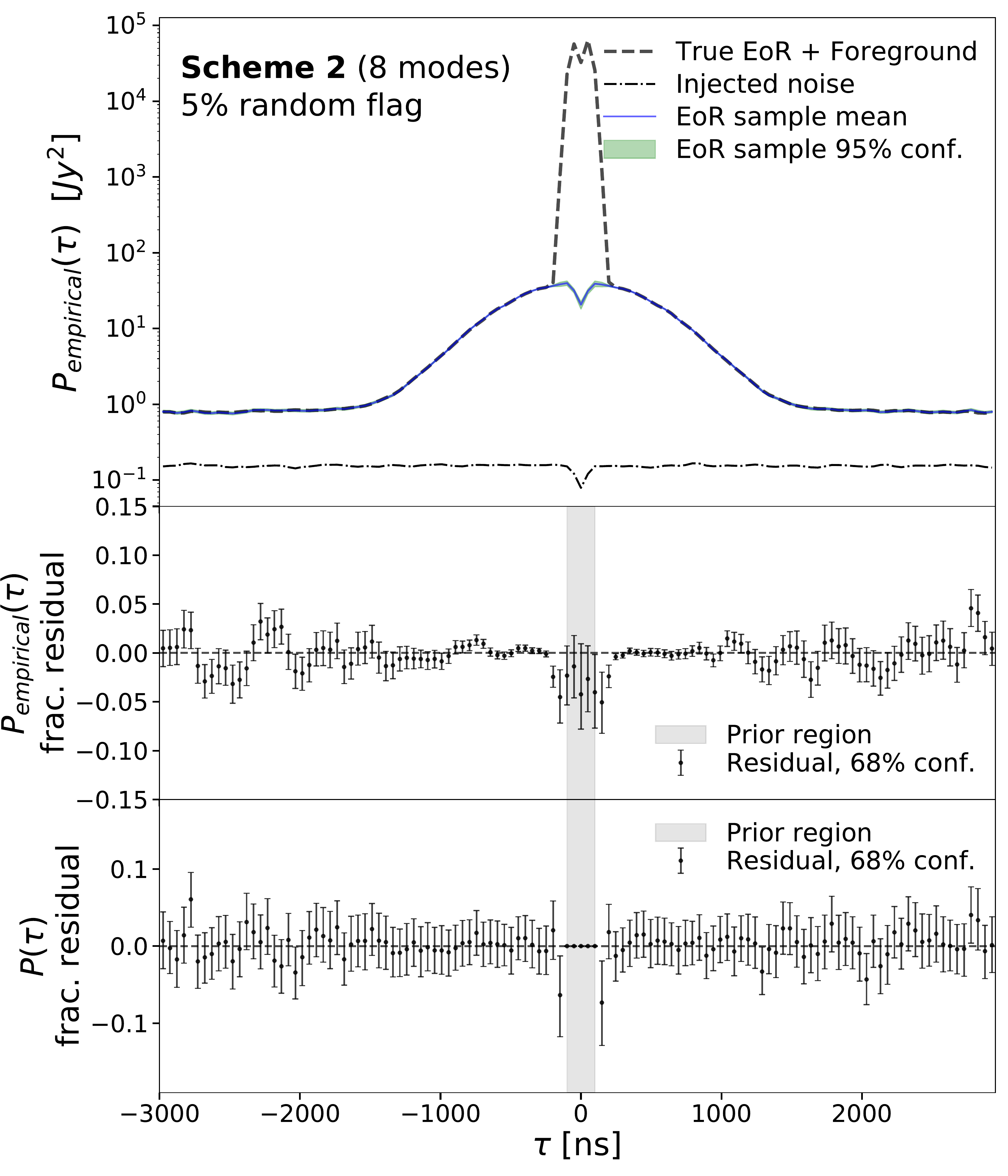}\hfill
	\includegraphics[width=\columnwidth]{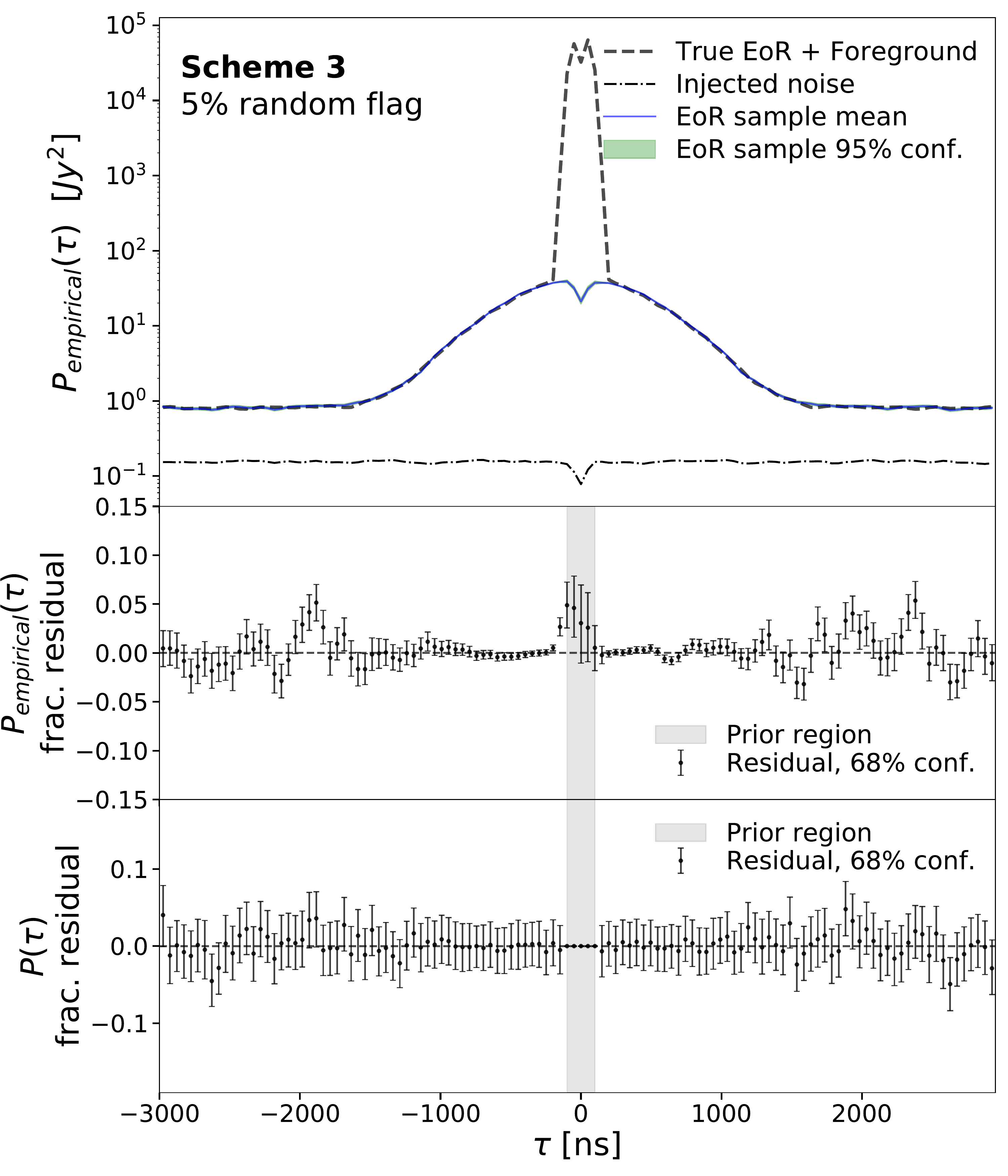}
	}
	\caption{Signal recovery using Scheme 2 (foreground template fitting, {\it left panel}) and Scheme 3 (joint foreground sampling, {\it right panel}). These runs match the parameters seen in Fig.~\ref{0-5-ci}; 800 total iterations, SNR at high delay of 5:1. These schemes differentiate between the foreground and EoR signal components and aim to recover the EoR component separately (unlike Scheme 1). {\it Top panel:} Mean of the EoR power spectrum estimated from the GCR samples of the EoR signal across 800 iterations, with 95$\%$ confidence region shown. The injected noise level is shown as a dash-dotted line.  {\it Middle panel:} Fractional residual between the true (input) EoR power spectrum and the mean of the power spectra derived from the GCR samples from the top panel. Outside the central delays where there is degeneracy between the foregrounds and EoR, the EoR signal recovery is essentially unbiased (within $\sim 5\%$ at high delay in both cases). Biases inside the degeneracy region are seen to take on slightly different shapes for Scheme 2 and Scheme 3. 68$\%$ confidence regions are also shown. {\it Bottom panel:} Fractional residual between the true delay spectrum and the inverse-Gamma sampled delay spectrum from each iteration, at the covariance sampling step of the sampling schemes. The zero residuals at the central delays are due to the prior that is imposed at low delays on this sampling step. Recovery of the delay spectrum is approximately unbiased when using both Gibbs sampling schemes under this test setup.}
	\label{5r-5-800}
\end{figure*}

\subsection{Comparison of the 3 Gibbs schemes} \label{sec:schemes}

Results from runs of each Gibbs sampling scheme under the conditions described above, for a 5\% random channel flagging pattern, are shown in Fig.~\ref{0-5-ci} (Scheme~1) and Fig.~\ref{5r-5-800} (Scheme 2 and 3). The top panel in each figure compares the true input power spectrum of the foregrounds and EoR signal, the injected noise level (representative of the noise on a visibility for a single time and baseline, i.e. before any time averaging) and the mean recovered EoR power spectrum (or total signal power spectrum, in the case of Fig.~\ref{0-5-ci}). When power spectra are calculated from samples of the frequency-space visibilities (i.e. in the top and second panels), the visibility data are first mean-subtracted and tapered with a Blackman-Harris window, since the foreground component of the visibilities is generally large and discontinuous at the band edges and so would otherwise cause ringing. This procedure causes a small dip at low delay in all power spectra. The second-from-the-top panels show the fractional residual between the true (input) signal power spectrum and the power spectrum estimated by squaring and averaging the GCR samples, along with 68$\%$ confidence intervals estimated from the GCR samples. Lower panels in Fig.~\ref{5r-5-800} show the recovered fractional residual between the true input signal power spectrum and the delay spectrum samples obtained at each iteration from the second (covariance sampling) step of the Gibbs sampling schemes.

\begin{figure*}
    {\centering
    \includegraphics[width=2\columnwidth]{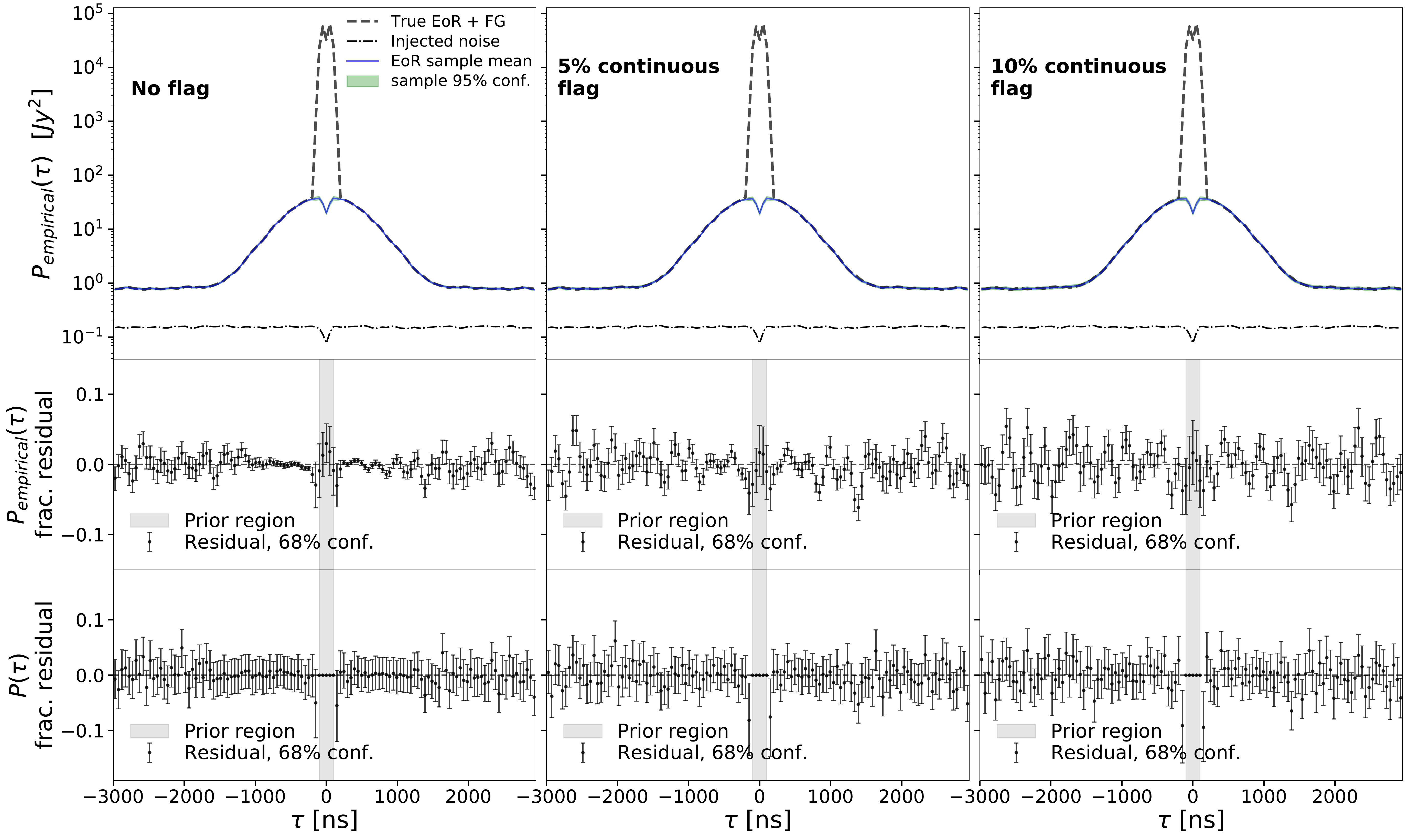}}
	\caption{Flagging fraction comparison using the Scheme 2 sampler (foreground templates) for 800 iterations, and with an SNR of 5:1 at high delay in each case. {\it Left column:} no flags applied. {\it Centre column:} Continuous flag through 5$\%$ of the band. {\it Right column:} Continuous flag through 10$\%$ of the band. The applied flagging is approximately in the centre of the frequency range and is applied to the data at all LSTs. Top panels show the mean $P_{\rm empirical}(\tau)$ and 95$\%$ confidence regions compared with the true power spectrum, and the injected noise level. Middle and lower panels show the fractional residual on $P_{\rm empirical}(\tau)$ and $P(\tau)$ samples across all iterations. Both statistics are approximately unbiased outside of the region of signal/foreground degeneracy ($|\tau|~<300\mu$s) in all flagging cases, though as the flagging fraction increases more fluctuations are noted in the residual means. Flagging can be seen to increase the width of the $P_{\rm empirical}(\tau)$ distribution, particularly at low delay. The width of the $P(\tau)$ residual does not appear to be altered significantly when increasing the flagging fraction, as this is dominated by sample variance, but the recovery of the `best-fit' delay spectrum (i.e. the central estimate shown with each errorbar) becomes noisier.}
	\label{flags-em-comparison}
\end{figure*}

Note that the delay spectrum estimate $P(\tau)$ obtained from the second step in the Gibbs scheme is in principle the correct one to use, as it properly takes into account sample variance. The power spectrum estimated from the GCR samples is the `empirical' power spectrum of the particular realisation of the EoR field that we see, and does not include sample variance. As such, it is expected to have narrower error bars, which is indeed what we see for Schemes 2 and 3.

In Schemes 2 and 3 shown in Fig.~\ref{5r-5-800} where the EoR signal is separated from the foreground component, the samplers exhibit a degeneracy in their solutions for the foreground power spectrum and EoR signal power spectrum where these components overlap in delay. Left unchecked, this degeneracy results in the sampler exploring the degeneracy region very slowly throughout the iterations. For this reason we implement a prior on the delay spectrum (covariance) samples at very low delay in these schemes, setting the delay spectrum samples in the 5 central delay bins to be equal to their true values, a `hard prior'. The region where the residual vanishes at low delay corresponds to this hard prior. This simplistic implementation of the prior is intended to show how these schemes can work when this degeneracy is prevented from being explored by the sampler. Other priors, such as a continuity prior, are likely to be appropriate in practice. We have checked that the results are not sensitive to the choice of hard prior; setting the prior on the delay spectrum in the central 5 bins to be 50$\%$ higher than the true value when carrying out the same run configurations, we found that neither the recovered signal power spectrum mean nor the errorbars were affected.

\begin{figure*}
    {\centering
    \includegraphics[width=2\columnwidth]{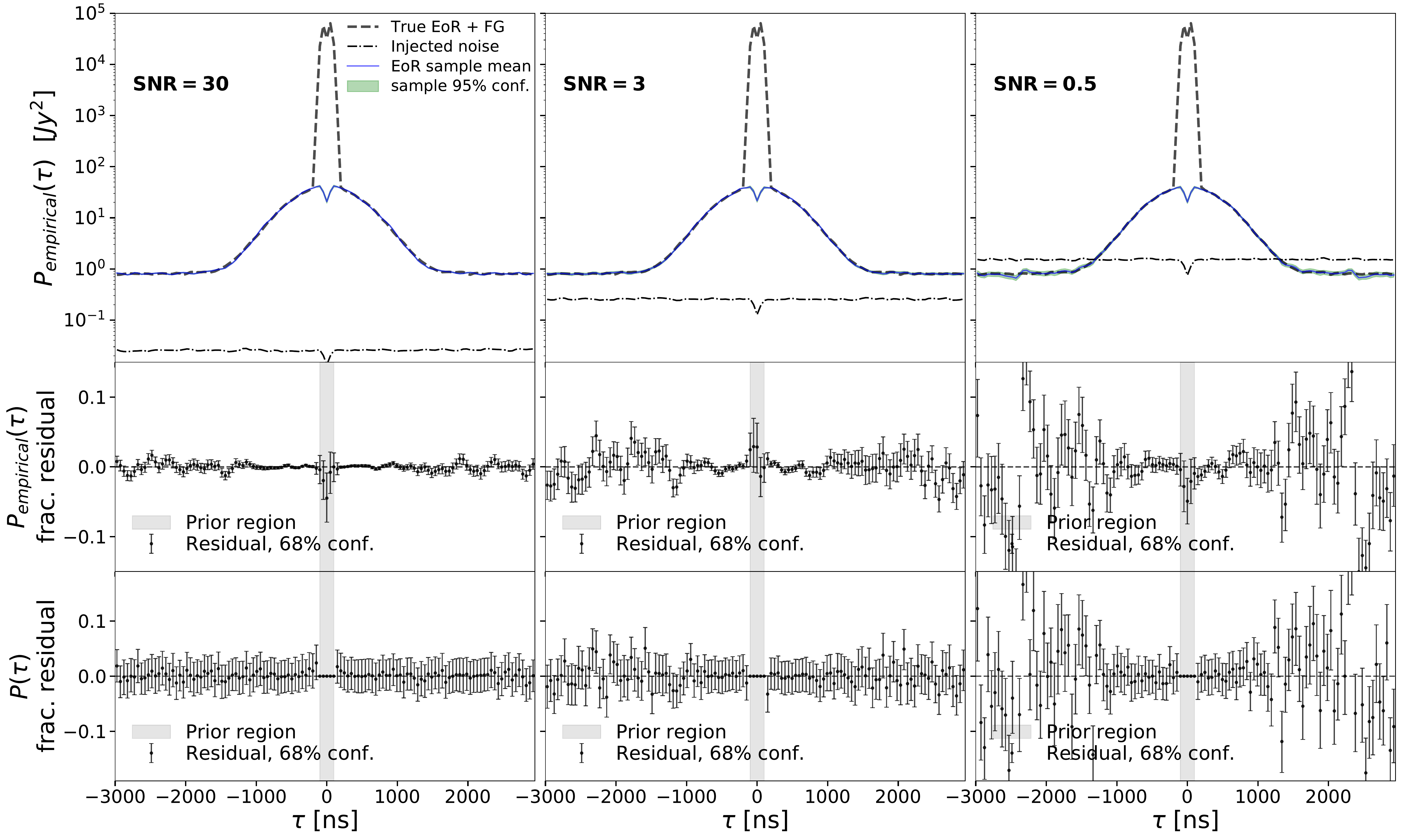}
	}
	\caption{SNR comparison using the Scheme 3 sampler (joint foreground sampling) for 800 iterations, and with no flags applied. {\it Left column:} SNR 30. {\it Centre column:} SNR 3. {\it Right column:} SNR 0.5. Top panels show the mean signal CR power spectrum and 95$\%$ confidence regions compared with the true power spectrum, and the injected noise level per LST (the other quantities are averaged over LST). Middle and lower panels show the fractional residual on $P_{\rm empirical}(\tau)$ and $P(\tau)$ across all iterations. As the SNR is decreased, larger fluctuations appear in the residual of the both statistics. The largest fluctuations in the residual appear at high delay, where the signal model is noise-like (flat) and the SNR is lowest.}
	
	\label{noise-jt-comparison}
\end{figure*}

The Scheme 1 `total signal' sampler in Fig.~\ref{0-5-ci} recovers a signal power spectrum that is biased low at higher delays, where the SNR is lowest. The reason for the bias is unclear, but we note that when the overall SNR is increased the bias not longer appears. Schemes 2 and 3 recover ostensibly unbiased signal power spectra as seen in their residuals (middle/lower panels), with the width of the residual distribution increasing towards higher delay where the SNR is lowest. Slight biases (dips) are noted for Scheme 2, inside the foreground delay range $|\tau|~<200$~ns, around the edge of the hard prior, where a degeneracy is expected between the signal and foreground amplitudes (see below). Immediately outside of this range, the middle panel shows a 68$\%$ confidence region spanning a fractional residual of approximately $\pm 0.5\%$, which increases to $\pm 2.5 \%$ at high delay, consistent with the SNR decreasing with delay. Wiggles with correlated errorbars are observed in the middle panels for both Schemes 2 and 3, which are expected as a taper has been applied. The delay spectrum samples in the lower panels also appear to be unbiased for both sampling schemes, with larger errorbars that show less evolution with delay, as sample variance is the dominant source of uncertainty. 

There is little to separate Scheme 2 and Scheme 3 based on their performance in the residual with the true input power spectrum, but Scheme 2, which relies on fitting a small number of foreground templates rather than the entire foreground covariance, reaches full sets of solutions for its GCR step approximately twice as fast as Scheme 3 does. We do however note the small dips in the recovered power spectra (both middle and lower panels) for Scheme 2 just outside the prior-dominated region of Fig.~\ref{5r-5-800}. We suspect that this is caused by the truncation of the set of foreground modes at the 8th mode. If this is correct, including more modes in the foreground model would allow more of the residual foreground emission at low delay to be absorbed, and this feature would not arise.

Finally, we note that an advantage of the Gibbs sampling approach is that the samples can be used to directly reconstruct the marginal posterior distributions of each parameter (or subset of parameters), without resorting to Gaussian approximations or otherwise. For the particular applications presented here, we did not find any particularly compelling examples of non-Gaussian behaviour of the marginal posteriors however. Visually inspecting the marginal distributions for the foreground amplitude parameters and delay spectrum bandpowers for Scheme 2, for the 10\% continuous flag and 10\% random flag cases, we found that they were generally consistent with Gaussianity, i.e. we did not note any strong skewness, heavy tails etc. More quantitative tests for Gaussianity could be performed if desired.

On a related note, the true EoR 21cm field is expected to have a non-Gaussian component (e.g. due to the formation of ionised bubbles around early sources), but we have modelled it as Gaussian in our analysis. Non-Gaussian features of the field can be captured in this framework, and the power spectrum is still a well-defined quantity that can be measured. Failing to explicitly account for the non-Gaussianity of the field will lead to these features, and therefore the power spectrum bandpowers, being weighted incorrectly however, and so statistics such as means and uncertainties could be biased. We leave an examination of this issue to future work.

We also calculated the covariance matrix for the parameters, finding no strong evidence for substantial covariance between the delay spectrum bandpowers and per-LST foreground amplitudes, even at low delay. This is likely due to a combination of the strong prior on the delay spectrum at low delay and the fact that the foreground amplitudes are estimated for each LST, whereas the bandpowers are estimated in an LST-averaged sense, which will tend to average down any correlations. We did however find covariance between the foreground amplitude parameters themselves. This is not unexpected considering that the foreground eigenmodes were calculated from the LST-averaged frequency-frequency covariance matrix, and so are not necessarily eigenmodes of an equivalent per-LST quantity (fits that use non-orthogonal basis functions will typically result in some covariance between their coefficients). We also noted a weak positive correlation between the $\tau = 0$~ns delay spectrum bandpower and other bandpowers, particularly at high delay, as well as a moderate correlation between the 1--2 delay spectrum bandpowers either side of the prior region. The latter is likely a manifestation of the incompleteness of the 8-mode foreground model, as mentioned above.

\begin{figure*}
    {\centering
	\includegraphics[width=\columnwidth]{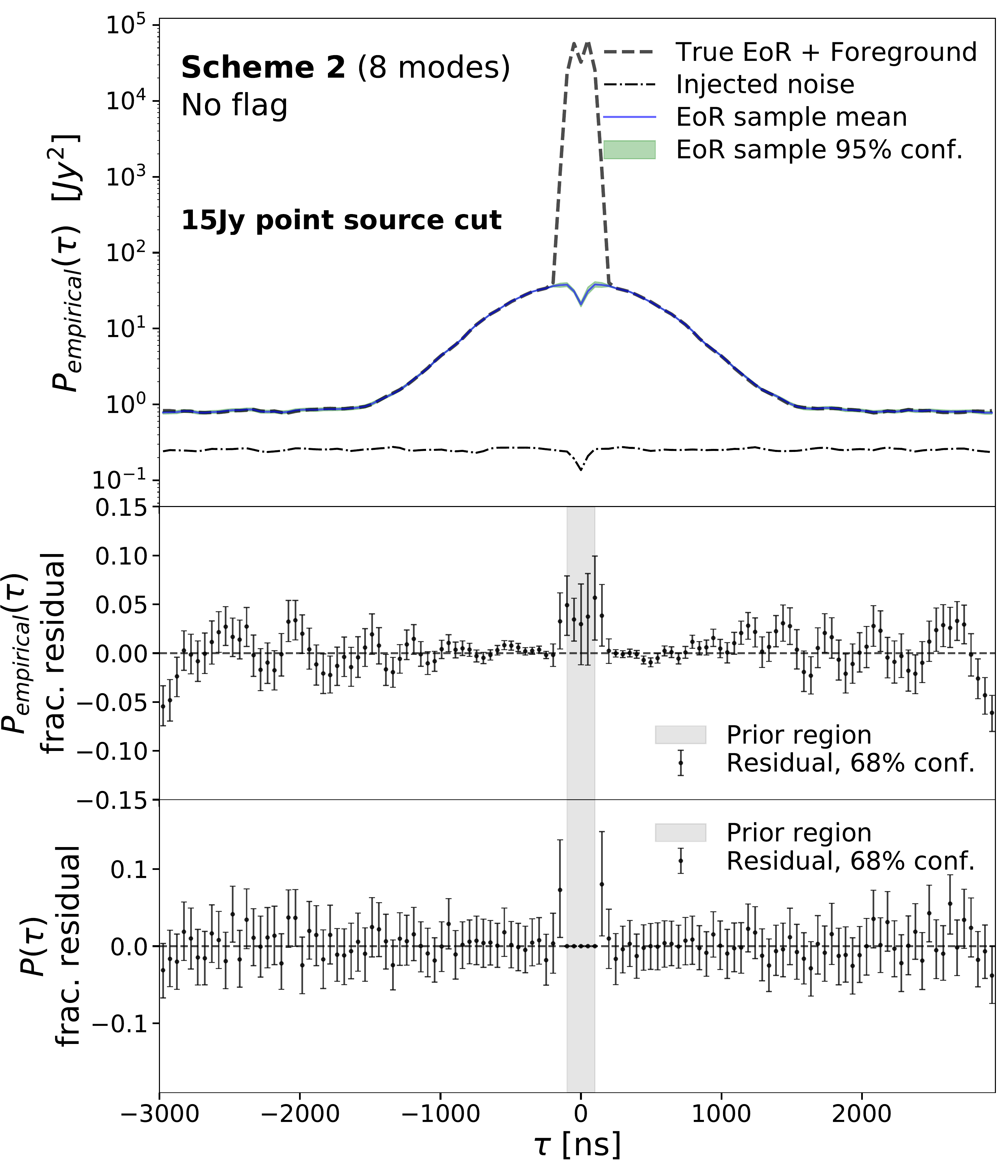}\hfill
	\includegraphics[width=\columnwidth]{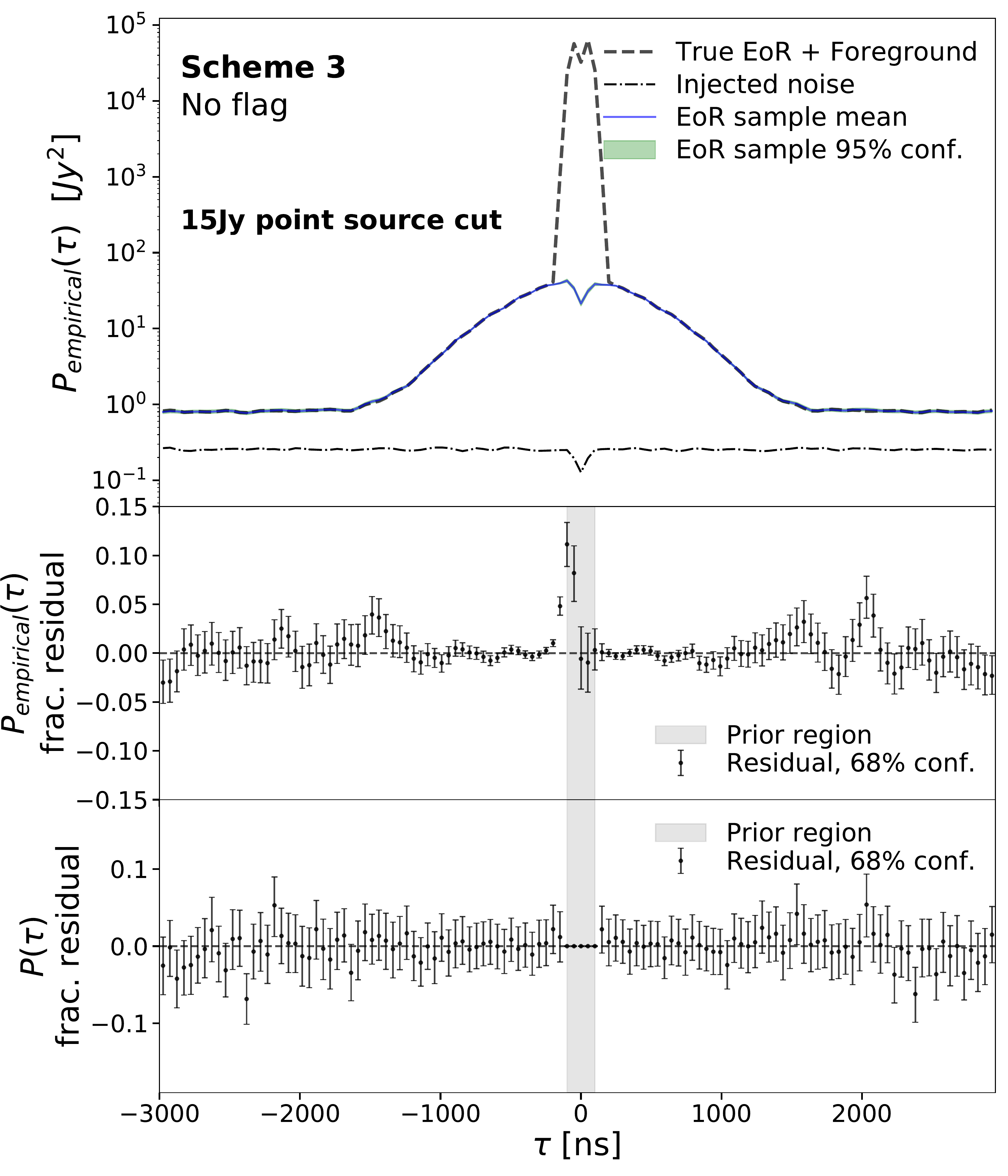}
	}
	\caption{Runs with Schemes 2 {\it(left panel)} and 3 {\it(right panel)} making use of foreground covariance matrix priors that have faint sources below 15~Jy cut from the catalogue. These runs are carried out to test the robustness of the method to faint source loss. The flux cut is set very   high in order to examine a reasonable worst-case scenario. Both runs return an unbiased residual, and have very similar errorbars to runs where the `correct' covariance matrix is used; e.g. the right panel is directly comparable with the center panels of Fig.~\ref{noise-jt-comparison}}
	\label{faint-comparison}
\end{figure*}

\vfill

\subsection{Dependence on flag fraction} \label{sec:flagfrac}

In Fig.~\ref{flags-em-comparison} we use Scheme 2 to assess how the signal power spectrum recovery changes when the flagging fraction is increased. In this study we use a continuous (rather than random) mask increasing up to 10$\%$ of the band in length, in order to evaluate a reasonable worst case scenario. We again add noise such that the SNR takes a value of 5 at high delay, and test a case with no flags, a 5$\%$ flag, and a 10$\%$ flag. The flags are applied near the centre of the frequency range and in the same position at all LSTs.

\begin{figure*}
    {\centering
	\includegraphics[width=1.05\columnwidth]{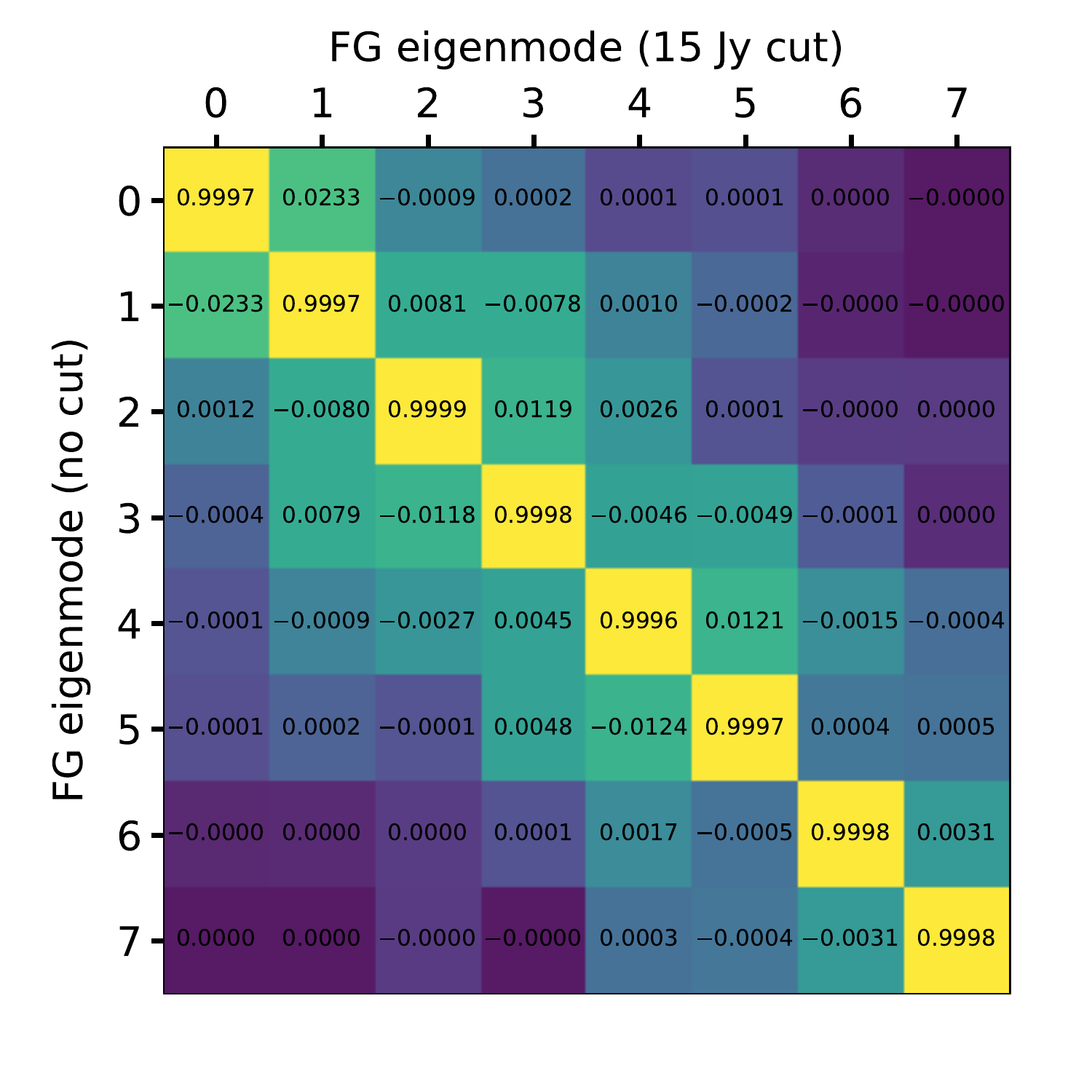}\hfill
	\includegraphics[width=1.05\columnwidth]{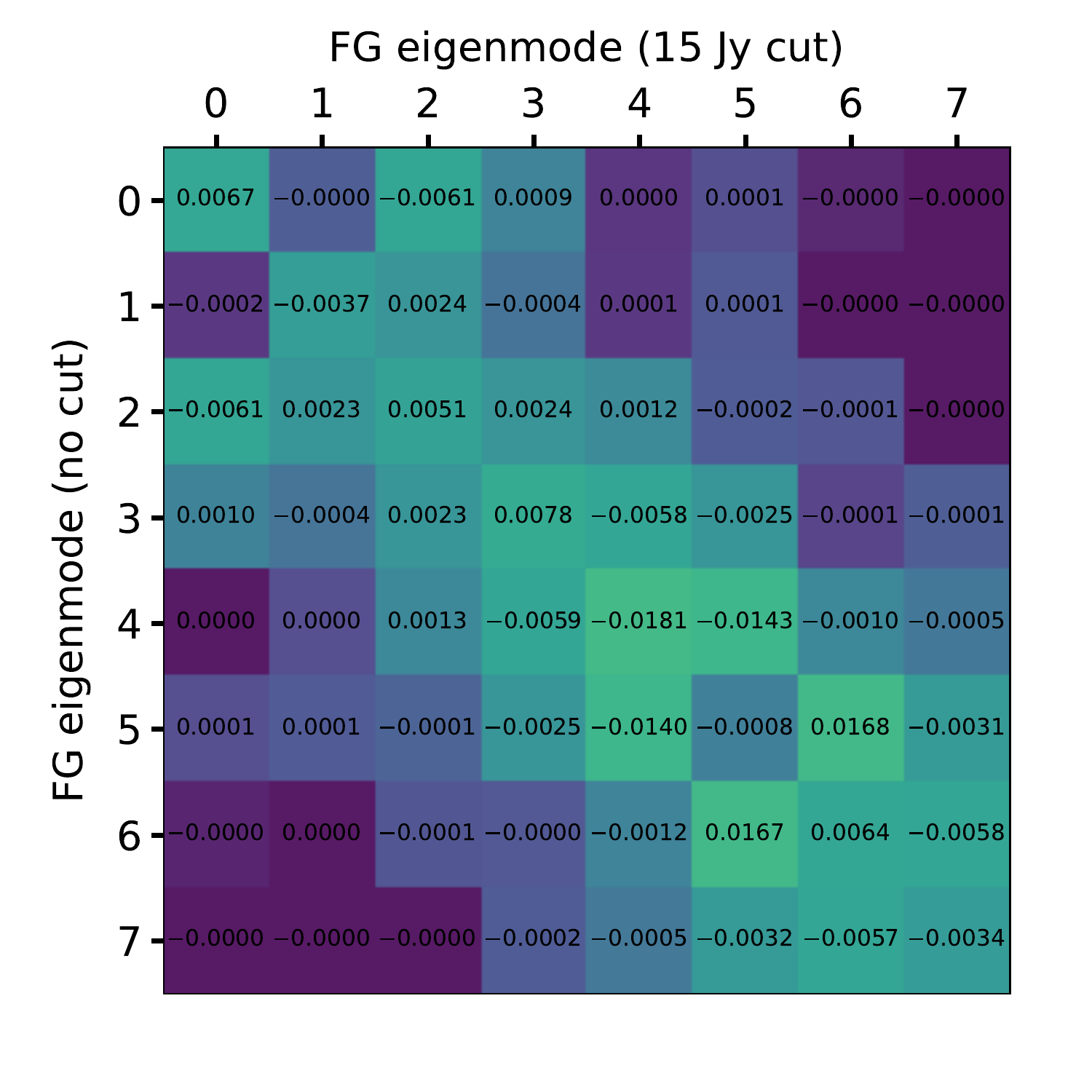}
	}
	\caption{Projection of the first 8 foreground eigenmodes for the `correct' foreground covariance matrix (vertical axis) onto the ones for the foreground model with a flux cut at 15~Jy (horizontal axis), calculated as $\bf{g}_{\rm full} \cdot \bf{g}_{\rm faint}^\dagger$. The real part is shown on the left and the imaginary part on the right. The colour scheme is chosen arbitrarily to highlight regions of similar value for the projection.}
	\label{fig:faint-eig-overlap}
\end{figure*}

The recovered signal power spectrum distributions in each of these runs again appear unbiased, even in the 10\% flagging case. The width of the errorbars increases as the size of the flagged region increases however, as one would expect from reducing the effective number of data points in the dataset. Looking at the middle panels, we see that the error bar width increases significantly at lower delays (e.g. just outside the hard prior region) as the flag fraction is increased.  
A large continuous flagging region in frequency space increases the overall uncertainty in all Fourier modes, but particularly prevents the lower-delay modes from being measured much more accurately that the higher delay modes, as they are in the no flagging case. Despite this, the power spectra sampled by the second Gibbs step remain essentially unchanged, as they are still dominated by sample variance. It is notable that no bias has been introduced into these power spectra either however; the samplers are successfully marginalising over the missing signal inside the region in such a way that the Gibbs-sampled power spectrum is recovered correctly. This is in contrast to (e.g.) a Wiener filter method, which would be biased towards zero signal (lower power) due to the flagged regions.

\subsection{Dependence on signal-to-noise ratio} \label{sec:snr}

Fig.~\ref{noise-jt-comparison} shows results from runs that use Scheme 3 to examine the effect of changing the SNR of the visibility data, with SNR values of 30, 3, and 0.5 at high delay being compared (reflected by the noise levels in the top panels). The data in these runs are unflagged to simplify the comparison. We change the SNR by rescaling the overall amplitudes of the EoR signal and foreground components, keeping the simulated noise level the same. 

The highest-SNR run obtains sub-percent error bars on the average signal power spectrum from the GCR step (middle panels). Decreasing the SNR by an order of magnitude (centre panels) causes larger fluctuations in the residual, on the order of 5$\%$, to appear in the recovered mean, along with an increase in the size of the errorbars (around 4$\%$ at high delay). At the lowest SNR of 0.5 (right panels), the size of the fluctuations in the residual approaches 15$\%$ at high delay, and an oscillation-like structure is observed, perhaps associated with the taper that has been applied. In the top panel of this run, the 95$\%$ confidence interval is much more clearly visible than it has been in any of the previous figures. This interval substantially increases in size once the EoR signal power falls below the noise level. Nevertheless, the recovered power spectra (in both the middle and lower panels) remain unbiased with respect to the true input power spectrum.

\subsection{Foreground covariance from an incomplete sky model} \label{sec:incomplete_fgcov}

Schemes 2 and 3 both require a model of the frequency-frequency covariance of the foregrounds. Any foreground model based on observations will inevitably be incomplete, as there will be potentially large numbers of faint sources missing from the source catalogue used to generate the model. We test the robustness of the Gibbs sampling schemes to this scenario. Fig.~\ref{faint-comparison} shows the results from runs using a foreground covariance matrix model that is generated from point source simulations with a lower source flux limit of 15~Jy, i.e. containing only the brightest sources. (The fluxes and spectral indices of the bright sources are assumed to be known and perfectly calibrated, however.) We carry out runs using Schemes 2 and 3 with a data SNR of 3 at high delay, and find that there is no qualitative difference with runs carried out using the correct foreground covariance matrix.

The `correct' (containing faint sources) foreground covariance matrix and the matrix generated without faint sources differ element-wise at sub-percent level, with the first 8 eigenmodes (as used by the Scheme 2 sampler) being close to identical on inspection. In Fig.~\ref{fig:faint-eig-overlap}, we show the scalar product of the first 8 eigenvectors of the correct foreground covariance matrix with the corresponding set of eigenvectors for the foreground model without faint sources. If the eigenvectors were identical, the real part of the plot (left panel) would be the identity matrix, and the imaginary part would be all zeros. As it is, the two sets of eigenvectors deviate from orthogonality at around the 1\% level or less.

Subject to the caveat that we have assumed a perfect model of the brightest sources, this result is encouraging: the frequency-frequency covariance matrix derived from a sky model containing all point sources, when averaged through the time direction, differs minimally from the same covariance matrix generated without faint sources. This is likely due to a few bright sources making a dominant contribution to the structure of the foreground templates. Depending on the baseline, there may be an ideal number of foreground templates (covariance eigenmodes) to use that capture the bright source contributions well, and change very minimally when the catalogue is incomplete. We have not studied the potential for optimisation further here however.

\section{Conclusions}\label{sec:conclusions}

Power spectrum estimation for 21cm EoR experiments is subject to a number of challenges, with particular difficulties arising from the large dynamic range between the EoR signal and foreground emission. Great care must be taken to avoid coupling Fourier modes inside and outside the foreground wedge, as otherwise leakage of the foregrounds into uncontaminated regions of Fourier space will occur. A serious source of leakage is ringing due to sharp edges caused by RFI masking; Fourier analysis becomes ill-posed in the case of a cut (partially-flagged) domain, and some way must be found to effectively model the contribution of the missing data to each Fourier mode. A common approach is to `in-paint' a plausible signal into the flagged regions to reduce the discontinuity with the unflagged data. Alternatively, knowing the flagging pattern and the covariance of the data, one could in principle `undo' most of the leakage by weighting the data by an appropriately masked inverse covariance matrix, which would serve to decorrelate the Fourier modes that were correlated by the masking. In both cases, the difficulty is in finding a model of the data and/or covariance matrix that is sufficiently accurate.

In this paper, we have presented an approach that effectively acts as a combined foreground separation, in-painting, and power spectrum estimation method. It is based on constructing a parametric model of the foregrounds, EoR signal, and their respective power spectra/covariance, and estimating the joint posterior distribution of all of these parameters using the combined techniques of Gibbs sampling and Gaussian constrained realisations (GCR). By exploiting the structure of the model and the Gaussianity of the likelihood, we can write the joint posterior of all the parameters as an iterative sampling scheme across multiple tractable conditional distributions for sub-spaces of the parameter space. With a Gaussian likelihood and suitable Gaussian or uniform priors, conditional distributions in which the model is linear in the parameters reduce to multivariate Gaussian distributions, which can be sampled from efficiently using GCR, a method based on a simple extension of Wiener filtering. This is possible even for very high dimensional spaces, potentially with hundreds of thousands of parameters. A useful byproduct of GCR (and Wiener filtering for that matter) is their ability to fill-in regions of missing data with plausible realisations of the data model, constrained by the surrounding unmasked data and an estimate of the signal covariance. A statistically well-posed in-painting step is therefore naturally incorporated in this method.

The iterative nature of the method also allows us to start with a relatively poor/simplistic estimate of the signal and foreground covariance matrices, but then converge to much better estimates through repeated sampling. Once convergence is achieved, the other sampling steps (e.g. the GCR steps) are (on average) using the true inverse covariance to weight the data, making it in some sense close to an `optimal' way of estimating the power spectra. This is not quite the same as performing an optimal quadratic estimate of the power spectrum however, as in that case one is conditioning on a fixed estimate of the data covariance, rather than marginalising over it. The marginalisation over the joint posterior of the signal and its covariance should also prevent signal loss occurring \citep[c.f.][]{kolopanis2019}, as possible correlations between the signal and the covariance are properly taken into account. We have gone some way towards a practical demonstration that this is the case in Sect.~\ref{sec:results}, in that we have shown that our methods are largely unbiased under Schemes 2 and 3, with Scheme 1 showing signal loss at a 5$\%$ level at high delay. We note that a similar effect can be achieved by marginalising over (rather than simply optimising) the kernel hyperparameters in Gaussian Process Regression (GPR) methods \citep{2021MNRAS.501.1463K}.

The Gibbs sampling approach permits a variety of modelling choices, particularly in terms of how to parametrise the foregrounds and the EoR covariance matrix/power spectrum. We have presented three Gibbs sampling schemes that sample from the joint posterior distribution of the EoR signal, its covariance, and some representation of the foreground emission. All three schemes make use of per-baseline foreground frequency-frequency covariance matrices estimated from point source simulations with similar properties (beams, etc.) to a subset of the HERA array.

We tested each of the samplers on a fiducial simulation containing 1200 LSTs for a single 14.6m East-West baseline, with a Gaussian EoR signal model with an SNR of 5 at high delay and 5$\%$ of the frequency band randomly flagged. We found that the `total signal' sampling scheme (Scheme 1), which samples the full signal-plus-foregrounds covariance matrix from a complex inverse Wishart distribution is biased at the 5$\%$ level at higher delays, but that this bias is not present when the SNR in the data is increased. This, along with the fact that sampling from a dense covariance matrix containing both foregrounds and EoR signal (as opposed to the EoR signal only, which can be modelled as diagonal in Fourier space) decreases the efficiency of exploring the joint posterior distribution, leads us to disfavour Scheme~1.

Conversely, we found that foreground template fitting (Scheme 2) and joint foreground sampling (Scheme 3) are able to recover unbiased signal power spectra when the degeneracy between EoR and foreground Fourier amplitudes is broken by a prior on the EoR signal at the lowest delays during the covariance sampling step. This would initially seem to be an overly strong prior assumption; however, we confirmed that using an incorrect hard prior (e.g. a 50$\%$ larger fixed value than the true EoR signal at low delay) does not influence EoR recovery at other delays. A different prior, such as a continuity prior, would be more appropriate in practice, and this would be an important line of inquiry for future work with Gibbs sampling schemes following this structure.

While our analysis has only considered a single 14.6m, E-W oriented baseline as an example, we anticipate similar behaviours of our method for longer baselines and different orientations, albeit with increasing spectral structure (and therefore more correlations with low-delay 21cm signal modes) as the baseline length increases, and as the orientation changes. The simulation-derived foreground frequency-frequency covariance matrices shown in Fig.~\ref{fgcovs} give some sense of how the correlation structure changes with baseline length, with the 43.8m baseline showing substantially reduced off-diagonal values compared with the 14.6m baselines for instance. Orientations away from the E-W direction also have a reduced level of frequency correlation, including enhanced off-diagonal components in the imaginary part. Studying the sensitivity of the method to these features, and the accuracy with which they are modelled, is left to future work.

We further tested the foreground template and joint foreground samplers under a range of flagging and noise conditions. We found that the width of the error bar on the recovered signal power spectrum increases when the flagging fraction is increased, particularly at lower delay, which is expected as more information has been lost by flagging a larger fraction of the data. We find that a lower SNR of the data also increases the uncertainty on the recovered EoR power spectrum, particularly when it falls below 1, again as expected.

Finally, we tested the robustness of the method to missing faint sources in the point source sky model used to estimate the foreground frequency-frequency covariance matrix. We found no qualitative difference in the recovery of the EoR power spectrum that was achieved, despite using quite a high flux cut of 15~Jy to model missing faint sources. We attribute this result to the fact that the covariance matrix of point source foregrounds (and particularly the leading principal components) changes comparatively little when faint sources are removed; the brightest sources appear to dominate its frequency structure. An important caveat is that we have ignored the contribution from diffuse foreground emission, which is likely to play an important role in determining the frequency structure of the foreground component, particularly for the shortest baselines.

Another caveat is that we have assumed that the EoR signal realisations are uncorrelated between time samples. This is an optimistic assumption, as one would only expect them to be independent on timescales of roughly the primary beam crossing time (i.e. long enough for a beam-sized patch of the sky to fully rotate through the primary beam mainlobe). Coherent averaging of visibilities on timescales of a few minutes would help a realistic analysis approach this idealisation, but it will still be important to quantify the time-time covariance of the signal and foregrounds in order to avoid overestimating the amount of independent information contained within a given dataset. The mathematical formalism required to incorporate time (as well as frequency) correlations should be a straightforward extension to what we have presented here, but the computational cost may be considerably higher.

Both the foreground template and joint foreground fitting schemes appear to be sufficient to recover the EoR power spectrum from individual baselines in an unbiased way, and with similar error properties.
In the absence of significant differences in recovery of the EoR, we note that the foreground template sampler (Scheme 2) completes iterations approximately twice as fast as the joint sampler (Scheme 3).

\section*{Acknowledgements}
We are grateful to J.~Burba, H.~Garsden, B.~Hazelton, N.~Kern, A.~Liu, and M.~Morales for useful comments and suggestions. This result is part of a project that has received funding from the European Research Council (ERC) under the European Union's Horizon 2020 research and innovation programme (Grant agreement No. 948764; PB, JB, and MJW). PB and FK acknowledge support from STFC Grants ST/T000341/1 and ST/X002624/1.

We acknowledge use of the following software: {\tt HEALPix} \citep{Gorski:2004by}, {\tt matplotlib} \citep{matplotlib}, {\tt numpy} \citep{numpy}, and {\tt scipy} \citep{2020SciPy-NMeth}.

\section*{Data Availability}
The code used to generate the results for this paper is available at \url{https://github.com/fraserlkennedy/Hydra-PSpec-prototype}.

\bibliographystyle{aasjournal}
\bibliography{references.bib} 

\begin{thebibliography}{}
\expandafter\ifx\csname natexlab\endcsname\relax\def\natexlab#1{#1}\fi
\providecommand{\url}[1]{\href{#1}{#1}}
\providecommand{\dodoi}[1]{doi:~\href{http://doi.org/#1}{\nolinkurl{#1}}}
\providecommand{\doeprint}[1]{\href{http://ascl.net/#1}{\nolinkurl{http://ascl.net/#1}}}
\providecommand{\doarXiv}[1]{\href{https://arxiv.org/abs/#1}{\nolinkurl{https://arxiv.org/abs/#1}}}

\bibitem[{{Ali} {et~al.}(2008){Ali}, {Bharadwaj}, \& {Chengalur}}]{ali08}
{Ali}, S.~S., {Bharadwaj}, S., \& {Chengalur}, J.~N. 2008, \mnras, 385, 2166,
  \dodoi{10.1111/j.1365-2966.2008.12984.x}

\bibitem[{Barry {et~al.}(2016)Barry, Hazelton, Sullivan, Morales, \&
  Pober}]{Barry:2016cpg}
Barry, N., Hazelton, B., Sullivan, I., Morales, M., \& Pober, J. 2016, Mon.
  Not. Roy. Astron. Soc., 461, 3135, \dodoi{10.1093/mnras/stw1380}

\bibitem[{{Barry} {et~al.}(2019){Barry}, {Wilensky}, {Trott}, {Pindor},
  {Beardsley}, {Hazelton}, {Sullivan}, {Morales}, {Pober}, {Line}, {Greig},
  {Byrne}, {Lanman}, {Li}, {Jordan}, {Joseph}, {McKinley}, {Rahimi},
  {Yoshiura}, {Bowman}, {Gaensler}, {Hewitt}, {Jacobs}, {Mitchell}, {Udaya
  Shankar}, {Sethi}, {Subrahmanyan}, {Tingay}, {Webster}, \&
  {Wyithe}}]{2019ApJ...884....1B}
{Barry}, N., {Wilensky}, M., {Trott}, C.~M., {et~al.} 2019, \apj, 884, 1,
  \dodoi{10.3847/1538-4357/ab40a8}

\bibitem[{{Bowman} {et~al.}(2009){Bowman}, {Morales}, \&
  {Hewitt}}]{2009ApJ...695..183B}
{Bowman}, J.~D., {Morales}, M.~F., \& {Hewitt}, J.~N. 2009, \apj, 695, 183,
  \dodoi{10.1088/0004-637X/695/1/183}

\bibitem[{{Byrne} {et~al.}(2019){Byrne}, {Morales}, {Hazelton}, {Li}, {Barry},
  {Beardsley}, {Joseph}, {Pober}, {Sullivan}, \& {Trott}}]{byrne19}
{Byrne}, R., {Morales}, M.~F., {Hazelton}, B., {et~al.} 2019, \apj, 875, 70,
  \dodoi{10.3847/1538-4357/ab107d}

\bibitem[{{Chakraborty} {et~al.}(2022){Chakraborty}, {Datta}, \&
  {Mazumder}}]{2022ApJ...929..104C}
{Chakraborty}, A., {Datta}, A., \& {Mazumder}, A. 2022, \apj, 929, 104,
  \dodoi{10.3847/1538-4357/ac5cc5}

\bibitem[{Chapman {et~al.}(2016)Chapman, Zaroubi, Abdalla, Dulwich, Jelić, \&
  Mort}]{10.1093/mnras/stw161}
Chapman, E., Zaroubi, S., Abdalla, F.~B., {et~al.} 2016, \mnras, 458, 2928,
  \dodoi{10.1093/mnras/stw161}

\bibitem[{{CHIME Collaboration} {et~al.}(2022){CHIME Collaboration}, {Amiri},
  {Bandura}, {Chen}, {Deng}, {Dobbs}, {Fandino}, {Foreman}, {Halpern}, {Hill},
  {Hinshaw}, {H{\"o}fer}, {Kania}, {Landecker}, {MacEachern}, {Masui},
  {Mena-Parra}, {Milutinovic}, {Mirhosseini}, {Newburgh}, {Ordog}, {Pen},
  {Pinsonneault-Marotte}, {Polzin}, {Reda}, {Renard}, {Shaw}, {Siegel},
  {Singh}, {Vanderlinde}, {Wang}, {Wiebe}, \& {Wulf}}]{2022arXiv220201242C}
{CHIME Collaboration}, {Amiri}, M., {Bandura}, K., {et~al.} 2022, arXiv
  e-prints, arXiv:2202.01242.
\newblock \doarXiv{2202.01242}

\bibitem[{{Choudhuri} {et~al.}(2021){Choudhuri}, {Bull}, \&
  {Garsden}}]{2021MNRAS.506.2066C}
{Choudhuri}, S., {Bull}, P., \& {Garsden}, H. 2021, \mnras, 506, 2066,
  \dodoi{10.1093/mnras/stab1795}

\bibitem[{{Datta} {et~al.}(2010){Datta}, {Bowman}, \&
  {Carilli}}]{2010ApJ...724..526D}
{Datta}, A., {Bowman}, J.~D., \& {Carilli}, C.~L. 2010, \apj, 724, 526,
  \dodoi{10.1088/0004-637X/724/1/526}

\bibitem[{{de Gasperin} {et~al.}(2019){de Gasperin}, {Dijkema}, {Drabent},
  {Mevius}, {Rafferty}, {van Weeren}, {Br{\"u}ggen}, {Callingham}, {Emig},
  {Heald}, {Intema}, {Morabito}, {Offringa}, {Oonk}, {Orr{\`u}},
  {R{\"o}ttgering}, {Sabater}, {Shimwell}, {Shulevski}, \&
  {Williams}}]{2019A&A...622A...5D}
{de Gasperin}, F., {Dijkema}, T.~J., {Drabent}, A., {et~al.} 2019, \aap, 622,
  A5, \dodoi{10.1051/0004-6361/201833867}

\bibitem[{{DeBoer} {et~al.}(2017){DeBoer}, {Parsons}, {Aguirre}, {Alexander},
  {Ali}, {Beardsley}, {Bernardi}, {Bowman}, {Bradley}, {Carilli}, {Cheng}, {de
  Lera Acedo}, {Dillon}, {Ewall-Wice}, {Fadana}, {Fagnoni}, {Fritz},
  {Furlanetto}, {Glendenning}, {Greig}, {Grobbelaar}, {Hazelton}, {Hewitt},
  {Hickish}, {Jacobs}, {Julius}, {Kariseb}, {Kohn}, {Lekalake}, {Liu}, {Loots},
  {MacMahon}, {Malan}, {Malgas}, {Maree}, {Martinot}, {Mathison}, {Matsetela},
  {Mesinger}, {Morales}, {Neben}, {Patra}, {Pieterse}, {Pober}, {Razavi-Ghods},
  {Ringuette}, {Robnett}, {Rosie}, {Sell}, {Smith}, {Syce}, {Tegmark},
  {Thyagarajan}, {Williams}, \& {Zheng}}]{2017PASP..129d5001D}
{DeBoer}, D.~R., {Parsons}, A.~R., {Aguirre}, J.~E., {et~al.} 2017, \pasp, 129,
  045001, \dodoi{10.1088/1538-3873/129/974/045001}

\bibitem[{{Di Matteo} {et~al.}(2002){Di Matteo}, {Perna}, {Abel}, \&
  {Rees}}]{2002ApJ...564..576D}
{Di Matteo}, T., {Perna}, R., {Abel}, T., \& {Rees}, M.~J. 2002, \apj, 564,
  576, \dodoi{10.1086/324293}

\bibitem[{{Eriksen} {et~al.}(2008){Eriksen}, {Jewell}, {Dickinson}, {Banday},
  {G{\'o}rski}, \& {Lawrence}}]{Eriksen}
{Eriksen}, H.~K., {Jewell}, J.~B., {Dickinson}, C., {et~al.} 2008, \apj, 676,
  10, \dodoi{10.1086/525277}

\bibitem[{{Ewall-Wice} {et~al.}(2017){Ewall-Wice}, {Dillon}, {Liu}, \&
  {Hewitt}}]{2017MNRAS.470.1849E}
{Ewall-Wice}, A., {Dillon}, J.~S., {Liu}, A., \& {Hewitt}, J. 2017, \mnras,
  470, 1849, \dodoi{10.1093/mnras/stx1221}

\bibitem[{{Ewall-Wice} {et~al.}(2016){Ewall-Wice}, {Dillon}, {Hewitt}, {Loeb},
  {Mesinger}, {Neben}, {Offringa}, {Tegmark}, {Barry}, {Beardsley}, {Bernardi},
  {Bowman}, {Briggs}, {Cappallo}, {Carroll}, {Corey}, {de Oliveira-Costa},
  {Emrich}, {Feng}, {Gaensler}, {Goeke}, {Greenhill}, {Hazelton},
  {Hurley-Walker}, {Johnston-Hollitt}, {Jacobs}, {Kaplan}, {Kasper}, {Kim},
  {Kratzenberg}, {Lenc}, {Line}, {Lonsdale}, {Lynch}, {McKinley}, {McWhirter},
  {Mitchell}, {Morales}, {Morgan}, {Thyagarajan}, {Oberoi}, {Ord}, {Paul},
  {Pindor}, {Pober}, {Prabu}, {Procopio}, {Riding}, {Rogers}, {Roshi},
  {Shankar}, {Sethi}, {Srivani}, {Subrahmanyan}, {Sullivan}, {Tingay}, {Trott},
  {Waterson}, {Wayth}, {Webster}, {Whitney}, {Williams}, {Williams}, {Wu}, \&
  {Wyithe}}]{2016MNRAS.460.4320E}
{Ewall-Wice}, A., {Dillon}, J.~S., {Hewitt}, J.~N., {et~al.} 2016, \mnras, 460,
  4320, \dodoi{10.1093/mnras/stw1022}

\bibitem[{{Ewall-Wice} {et~al.}(2021){Ewall-Wice}, {Kern}, {Dillon}, {Liu},
  {Parsons}, {Singh}, {Lanman}, {La Plante}, {Fagnoni}, {Acedo}, {DeBoer},
  {Nunhokee}, {Bull}, {Chang}, {Lazio}, {Aguirre}, \&
  {Weinberg}}]{2021MNRAS.500.5195E}
{Ewall-Wice}, A., {Kern}, N., {Dillon}, J.~S., {et~al.} 2021, \mnras, 500,
  5195, \dodoi{10.1093/mnras/staa3293}

\bibitem[{{Fagnoni} {et~al.}(2021){Fagnoni}, {de Lera Acedo}, {DeBoer},
  {Abdurashidova}, {Aguirre}, {Alexander}, {Ali}, {Balfour}, {Beardsley},
  {Bernardi}, {Billings}, {Bowman}, {Bradley}, {Bull}, {Burba}, {Carilli},
  {Cheng}, {Dexter}, {Dillon}, {Ewall-Wice}, {Fritz}, {Furlanetto},
  {Gale-Sides}, {Glendenning}, {Gorthi}, {Greig}, {Grobbelaar}, {Halday},
  {Hazelton}, {Hewitt}, {Hickish}, {Jacobs}, {Josaitis}, {Julius}, {Kern},
  {Kerrigan}, {Kim}, {Kittiwisit}, {Kohn}, {Kolopanis}, {Lanman}, {Plante},
  {Lekalake}, {Liu}, {MacMahon}, {Malan}, {Malgas}, {Maree}, {Martinot},
  {Matsetela}, {Mena Parra}, {Mesinger}, {Molewa}, {Morales}, {Mosiane},
  {Neben}, {Nikolic}, {Parsons}, {Patra}, {Pieterse}, {Pober}, {Razavi-Ghods},
  {Robnett}, {Rosie}, {Sims}, {Smith}, {Syce}, {Thyagarajan}, {Williams}, \&
  {Zheng}}]{2021MNRAS.500.1232F}
{Fagnoni}, N., {de Lera Acedo}, E., {DeBoer}, D.~R., {et~al.} 2021, \mnras,
  500, 1232, \dodoi{10.1093/mnras/staa3268}

\bibitem[{{Furlanetto} {et~al.}(2006){Furlanetto}, {Oh}, \&
  {Briggs}}]{2006PhR...433..181F}
{Furlanetto}, S.~R., {Oh}, S.~P., \& {Briggs}, F.~H. 2006, \physrep, 433, 181,
  \dodoi{10.1016/j.physrep.2006.08.002}

\bibitem[{Gallager(2013)}]{Gallager2013}
Gallager, R. 2013, Stochastic Processes: Theory for Applications (CUP),
  \dodoi{10.1017/CBO9781139626514}

\bibitem[{{Garsden} {et~al.}(2021){Garsden}, {Greenhill}, {Bernardi},
  {Fialkov}, {Price}, {Mitchell}, {Dowell}, {Spinelli}, \&
  {Schinzel}}]{2021MNRAS.506.5802G}
{Garsden}, H., {Greenhill}, L., {Bernardi}, G., {et~al.} 2021, \mnras, 506,
  5802, \dodoi{10.1093/mnras/stab1671}

\bibitem[{{Gehlot} {et~al.}(2018){Gehlot}, {Koopmans}, {de Bruyn}, {Zaroubi},
  {Brentjens}, {Asad}, {Hatef}, {Jeli{\'c}}, {Mevius}, {Offringa}, {Pandey}, \&
  {Yatawatta}}]{2018MNRAS.478.1484G}
{Gehlot}, B.~K., {Koopmans}, L.~V.~E., {de Bruyn}, A.~G., {et~al.} 2018,
  \mnras, 478, 1484, \dodoi{10.1093/mnras/sty1095}

\bibitem[{Geman \& Geman(1984)}]{geman1984stochastic}
Geman, S., \& Geman, D. 1984, IEEE Transactions on pattern analysis and machine
  intelligence, 721

\bibitem[{{Ghosh} {et~al.}(2020){Ghosh}, {Mertens}, {Bernardi}, {Santos},
  {Kern}, {Carilli}, {Grobler}, {Koopmans}, {Jacobs}, {Liu}, {Parsons},
  {Morales}, {Aguirre}, {Dillon}, {Hazelton}, {Smirnov}, {Gehlot}, {Matika},
  {Alexander}, {Ali}, {Beardsley}, {Benefo}, {Billings}, {Bowman}, {Bradley},
  {Cheng}, {Chichura}, {DeBoer}, {de Lera Acedo}, {Ewall-Wice}, {Fadana},
  {Fagnoni}, {Fortino}, {Fritz}, {Furlanetto}, {Gallardo}, {Glendenning},
  {Gorthi}, {Greig}, {Grobbelaar}, {Hickish}, {Josaitis}, {Julius}, {Igarashi},
  {Kariseb}, {Kohn}, {Kolopanis}, {Lekalake}, {Loots}, {MacMahon}, {Malan},
  {Malgas}, {Maree}, {Martinot}, {Mathison}, {Matsetela}, {Mesinger}, {Neben},
  {Nikolic}, {Nunhokee}, {Patra}, {Pieterse}, {Razavi-Ghods}, {Ringuette},
  {Robnett}, {Rosie}, {Sell}, {Smith}, {Syce}, {Tegmark}, {Thyagarajan},
  {Williams}, \& {Zheng}}]{2020MNRAS.495.2813G}
{Ghosh}, A., {Mertens}, F., {Bernardi}, G., {et~al.} 2020, \mnras, 495, 2813,
  \dodoi{10.1093/mnras/staa1331}

\bibitem[{Goodman(1963)}]{goodman1963}
Goodman, N.~R. 1963, The Annals of Mathematical Statistics, 34, 152 ,
  \dodoi{10.1214/aoms/1177704250}

\bibitem[{Gorski {et~al.}(2005)Gorski, Hivon, Banday, Wandelt, Hansen,
  Reinecke, \& Bartelman}]{Gorski:2004by}
Gorski, K.~M., Hivon, E., Banday, A.~J., {et~al.} 2005, Astrophys. J., 622,
  759, \dodoi{10.1086/427976}

\bibitem[{{Hallinan}(2014)}]{2014era..conf10203H}
{Hallinan}, G. 2014, in Exascale Radio Astronomy, Vol.~2, 10203

\bibitem[{{Hunter}(2007)}]{matplotlib}
{Hunter}, J.~D. 2007, Computing in Science Engineering, 9, 90

\bibitem[{{Hurley-Walker} {et~al.}(2017){Hurley-Walker}, {Callingham},
  {Hancock}, {Franzen}, {Hindson}, {Kapi{\'n}ska}, {Morgan}, {Offringa},
  {Wayth}, {Wu}, {Zheng}, {Murphy}, {Bell}, {Dwarakanath}, {For}, {Gaensler},
  {Johnston-Hollitt}, {Lenc}, {Procopio}, {Staveley-Smith}, {Ekers}, {Bowman},
  {Briggs}, {Cappallo}, {Deshpande}, {Greenhill}, {Hazelton}, {Kaplan},
  {Lonsdale}, {McWhirter}, {Mitchell}, {Morales}, {Morgan}, {Oberoi}, {Ord},
  {Prabu}, {Shankar}, {Srivani}, {Subrahmanyan}, {Tingay}, {Webster},
  {Williams}, \& {Williams}}]{hurley17}
{Hurley-Walker}, N., {Callingham}, J.~R., {Hancock}, P.~J., {et~al.} 2017,
  \mnras, 464, 1146, \dodoi{10.1093/mnras/stw2337}

\bibitem[{{Joseph} {et~al.}(2018){Joseph}, {Trott}, \& {Wayth}}]{joseph18}
{Joseph}, R.~C., {Trott}, C.~M., \& {Wayth}, R.~B. 2018, \aj, 156, 285,
  \dodoi{10.3847/1538-3881/aaec0b}

\bibitem[{{Joseph} {et~al.}(2020){Joseph}, {Trott}, {Wayth}, \&
  {Nasirudin}}]{joseph20}
{Joseph}, R.~C., {Trott}, C.~M., {Wayth}, R.~B., \& {Nasirudin}, A. 2020,
  \mnras, 492, 2017, \dodoi{10.1093/mnras/stz3375}

\bibitem[{{Kariuki Chege} {et~al.}(2022){Kariuki Chege}, {Jordan}, {Lynch},
  {Trott}, {Line}, {Pindor}, \& {Yoshiura}}]{2022arXiv220712090K}
{Kariuki Chege}, J., {Jordan}, C.~H., {Lynch}, C., {et~al.} 2022, arXiv
  e-prints, arXiv:2207.12090, \dodoi{10.48550/arXiv.2207.12090}

\bibitem[{{Kern} \& {Liu}(2021)}]{2021MNRAS.501.1463K}
{Kern}, N.~S., \& {Liu}, A. 2021, \mnras, 501, 1463,
  \dodoi{10.1093/mnras/staa3736}

\bibitem[{{Kern} {et~al.}(2019){Kern}, {Parsons}, {Dillon}, {Lanman},
  {Fagnoni}, \& {de Lera Acedo}}]{2019ApJ...884..105K}
{Kern}, N.~S., {Parsons}, A.~R., {Dillon}, J.~S., {et~al.} 2019, \apj, 884,
  105, \dodoi{10.3847/1538-4357/ab3e73}

\bibitem[{{Kolopanis} {et~al.}(2019){Kolopanis}, {Jacobs}, {Cheng}, {Parsons},
  {Kohn}, {Pober}, {Aguirre}, {Ali}, {Bernardi}, {Bradley}, {Carilli},
  {DeBoer}, {Dexter}, {Dillon}, {Kerrigan}, {Klima}, {Liu}, {MacMahon},
  {Moore}, {Thyagarajan}, {Nunhokee}, {Walbrugh}, \& {Walker}}]{kolopanis2019}
{Kolopanis}, M., {Jacobs}, D.~C., {Cheng}, C., {et~al.} 2019, \apj, 883, 133,
  \dodoi{10.3847/1538-4357/ab3e3a}

\bibitem[{{Li} {et~al.}(2019){Li}, {Pober}, {Barry}, {Hazelton}, {Morales},
  {Trott}, {Lanman}, {Wilensky}, {Sullivan}, {Beardsley}, {Booler}, {Bowman},
  {Byrne}, {Crosse}, {Emrich}, {Franzen}, {Hasegawa}, {Horsley},
  {Johnston-Hollitt}, {Jacobs}, {Jordan}, {Joseph}, {Kaneuji}, {Kaplan},
  {Kenney}, {Kubota}, {Line}, {Lynch}, {McKinley}, {Mitchell}, {Murray},
  {Pallot}, {Pindor}, {Rahimi}, {Riding}, {Sleap}, {Steele}, {Takahashi},
  {Tingay}, {Walker}, {Wayth}, {Webster}, {Williams}, {Wu}, {Wyithe},
  {Yoshiura}, \& {Zheng}}]{2019ApJ...887..141L}
{Li}, W., {Pober}, J.~C., {Barry}, N., {et~al.} 2019, \apj, 887, 141,
  \dodoi{10.3847/1538-4357/ab55e4}

\bibitem[{{Liu} {et~al.}(2014{\natexlab{a}}){Liu}, {Parsons}, \&
  {Trott}}]{2014PhRvD..90b3018L}
{Liu}, A., {Parsons}, A.~R., \& {Trott}, C.~M. 2014{\natexlab{a}}, \prd, 90,
  023018, \dodoi{10.1103/PhysRevD.90.023018}

\bibitem[{{Liu} {et~al.}(2014{\natexlab{b}}){Liu}, {Parsons}, \&
  {Trott}}]{2014PhRvD..90b3019L}
---. 2014{\natexlab{b}}, \prd, 90, 023019, \dodoi{10.1103/PhysRevD.90.023019}

\bibitem[{{Madau} {et~al.}(1997){Madau}, {Meiksin}, \&
  {Rees}}]{1997ApJ...475..429M}
{Madau}, P., {Meiksin}, A., \& {Rees}, M.~J. 1997, \apj, 475, 429,
  \dodoi{10.1086/303549}

\bibitem[{{Mertens} {et~al.}(2020){Mertens}, {Mevius}, {Koopmans}, {Offringa},
  {Mellema}, {Zaroubi}, {Brentjens}, {Gan}, {Gehlot}, {Pandey}, {Sardarabadi},
  {Vedantham}, {Yatawatta}, {Asad}, {Ciardi}, {Chapman}, {Gazagnes}, {Ghara},
  {Ghosh}, {Giri}, {Iliev}, {Jeli{\'c}}, {Kooistra}, {Mondal}, {Schaye}, \&
  {Silva}}]{2020MNRAS.493.1662M}
{Mertens}, F.~G., {Mevius}, M., {Koopmans}, L.~V.~E., {et~al.} 2020, \mnras,
  493, 1662, \dodoi{10.1093/mnras/staa327}

\bibitem[{Messerschmitt(2006)}]{messerschmitt2006}
Messerschmitt, D. 2006, EECS Dept., Univ. of California, Berkeley, Tech. Rep.
  No. UCB/EECS-2006-90

\bibitem[{{Morales} {et~al.}(2006){Morales}, {Bowman}, \&
  {Hewitt}}]{2006ApJ...648..767M}
{Morales}, M.~F., {Bowman}, J.~D., \& {Hewitt}, J.~N. 2006, \apj, 648, 767,
  \dodoi{10.1086/506135}

\bibitem[{{Morales} {et~al.}(2012){Morales}, {Hazelton}, {Sullivan}, \&
  {Beardsley}}]{2012ApJ...752..137M}
{Morales}, M.~F., {Hazelton}, B., {Sullivan}, I., \& {Beardsley}, A. 2012,
  \apj, 752, 137, \dodoi{10.1088/0004-637X/752/2/137}

\bibitem[{{Morales} \& {Wyithe}(2010)}]{morales10}
{Morales}, M.~F., \& {Wyithe}, J. S.~B. 2010, \araa, 48, 127,
  \dodoi{10.1146/annurev-astro-081309-130936}

\bibitem[{{Mouri Sardarabadi} \& {Koopmans}(2019)}]{2019MNRAS.483.5480M}
{Mouri Sardarabadi}, A., \& {Koopmans}, L.~V.~E. 2019, \mnras, 483, 5480,
  \dodoi{10.1093/mnras/sty3444}

\bibitem[{Murray {et~al.}(2017)Murray, Trott, \& Jordan}]{murray2017}
Murray, S., Trott, C., \& Jordan, C. 2017, Proceedings of the International
  Astronomical Union, 12, 199

\bibitem[{{Offringa} {et~al.}(2019){Offringa}, {Mertens}, \&
  {Koopmans}}]{2019MNRAS.484.2866O}
{Offringa}, A.~R., {Mertens}, F., \& {Koopmans}, L.~V.~E. 2019, \mnras, 484,
  2866, \dodoi{10.1093/mnras/stz175}

\bibitem[{Orosz {et~al.}(2019)Orosz, Dillon, Ewall-Wice, Parsons, \&
  Thyagarajan}]{Orosz:2018avj}
Orosz, N., Dillon, J.~S., Ewall-Wice, A., Parsons, A.~R., \& Thyagarajan, N.
  2019, MNRAS, 487, 537, \dodoi{10.1093/mnras/stz1287}

\bibitem[{Paciga {et~al.}(2013)Paciga, Albert, Bandura, Chang, Gupta, Hirata,
  Odegova, Pen, Peterson, Roy, Shaw, Sigurdson, \& Voytek}]{Paciga2013}
Paciga, G., Albert, J.~G., Bandura, K., {et~al.} 2013, Monthly Notices of the
  Royal Astronomical Society, 433, 639, \dodoi{10.1093/mnras/stt753}

\bibitem[{{Pagano} {et~al.}(2022){Pagano}, {Liu}, {Liu}, {Kern}, {Ewall-Wice},
  {Bull}, {Pascua}, {Ravanbakhsh}, {Abdurashidova}, {Adams}, {Aguirre},
  {Alexander}, {Ali}, {Baartman}, {Balfour}, {Beardsley}, {Bernardi},
  {Billings}, {Bowman}, {Bradley}, {Burba}, {Carey}, {Carilli}, {Cheng},
  {DeBoer}, {de Lera Acedo}, {Dexter}, {Dillon}, {Eksteen}, {Ely}, {Fagnoni},
  {Fritz}, {Furlanetto}, {Gale-Sides}, {Glendenning}, {Gorthi}, {Greig},
  {Grobbelaar}, {Halday}, {Hazelton}, {Hewitt}, {Hickish}, {Jacobs}, {Julius},
  {Kariseb}, {Kerrigan}, {Kittiwisit}, {Kohn}, {Kolopanis}, {Lanman}, {La
  Plante}, {Loots}, {MacMahon}, {Malan}, {Malgas}, {Malgas}, {Marero},
  {Martinot}, {Mesinger}, {Molewa}, {Morales}, {Mosiane}, {Neben}, {Nikolic},
  {Nuwegeld}, {Parsons}, {Patra}, {Pieterse}, {Razavi-Ghods}, {Robnett},
  {Rosie}, {Sims}, {Smith}, {Swarts}, {Thyagarajan}, {van Wyngaarden},
  {Williams}, \& {Zheng}}]{Pagano2022}
{Pagano}, M., {Liu}, J., {Liu}, A., {et~al.} 2022, arXiv e-prints,
  arXiv:2210.14927.
\newblock \doarXiv{2210.14927}

\bibitem[{{Parsons} \& {Backer}(2009)}]{2009AJ....138..219P}
{Parsons}, A.~R., \& {Backer}, D.~C. 2009, \aj, 138, 219,
  \dodoi{10.1088/0004-6256/138/1/219}

\bibitem[{{Parsons} {et~al.}(2010){Parsons}, {Backer}, {Foster}, {Wright},
  {Bradley}, {Gugliucci}, {Parashare}, {Benoit}, {Aguirre}, {Jacobs},
  {Carilli}, {Herne}, {Lynch}, {Manley}, \& {Werthimer}}]{2010AJ....139.1468P}
{Parsons}, A.~R., {Backer}, D.~C., {Foster}, G.~S., {et~al.} 2010, \aj, 139,
  1468, \dodoi{10.1088/0004-6256/139/4/1468}

\bibitem[{{Patil} {et~al.}(2016){Patil}, {Yatawatta}, {Zaroubi}, {Koopmans},
  {de Bruyn}, {Jeli{\'c}}, {Ciardi}, {Iliev}, {Mevius}, {Pandey}, \&
  {Gehlot}}]{2016MNRAS.463.4317P}
{Patil}, A.~H., {Yatawatta}, S., {Zaroubi}, S., {et~al.} 2016, \mnras, 463,
  4317, \dodoi{10.1093/mnras/stw2277}

\bibitem[{{Patil} {et~al.}(2017){Patil}, {Yatawatta}, {Koopmans}, {de Bruyn},
  {Brentjens}, {Zaroubi}, {Asad}, {Hatef}, {Jeli{\'c}}, {Mevius}, {Offringa},
  {Pandey}, {Vedantham}, {Abdalla}, {Brouw}, {Chapman}, {Ciardi}, {Gehlot},
  {Ghosh}, {Harker}, {Iliev}, {Kakiichi}, {Majumdar}, {Mellema}, {Silva},
  {Schaye}, {Vrbanec}, \& {Wijnholds}}]{2017ApJ...838...65P}
{Patil}, A.~H., {Yatawatta}, S., {Koopmans}, L.~V.~E., {et~al.} 2017, \apj,
  838, 65, \dodoi{10.3847/1538-4357/aa63e7}

\bibitem[{{Pober} {et~al.}(2013){Pober}, {Parsons}, {Aguirre}, {Ali},
  {Bradley}, {Carilli}, {DeBoer}, {Dexter}, {Gugliucci}, {Jacobs}, {Klima},
  {MacMahon}, {Manley}, {Moore}, {Stefan}, \& {Walbrugh}}]{2013ApJ...768L..36P}
{Pober}, J.~C., {Parsons}, A.~R., {Aguirre}, J.~E., {et~al.} 2013, \apjl, 768,
  L36, \dodoi{10.1088/2041-8205/768/2/L36}

\bibitem[{{Pritchard} \& {Loeb}(2012)}]{2012RPPh...75h6901P}
{Pritchard}, J.~R., \& {Loeb}, A. 2012, Reports on Progress in Physics, 75,
  086901, \dodoi{10.1088/0034-4885/75/8/086901}

\bibitem[{{Rybicki} \& {Press}(1992)}]{1992ApJ...398..169R}
{Rybicki}, G.~B., \& {Press}, W.~H. 1992, \apj, 398, 169,
  \dodoi{10.1086/171845}

\bibitem[{{Santos} {et~al.}(2005){Santos}, {Cooray}, \& {Knox}}]{santos05}
{Santos}, M.~G., {Cooray}, A., \& {Knox}, L. 2005, \apj, 625, 575,
  \dodoi{10.1086/429857}

\bibitem[{S{\"a}rkk{\"a} \& Solin(2013)}]{sarkka2013continuous}
S{\"a}rkk{\"a}, S., \& Solin, A. 2013, in Scandinavian Conference on Image
  Analysis, Springer, 172--181

\bibitem[{{Swarup} {et~al.}(1991){Swarup}, {Ananthakrishnan}, {Kapahi}, {Rao},
  {Subrahmanya}, \& {Kulkarni}}]{swarup91}
{Swarup}, G., {Ananthakrishnan}, S., {Kapahi}, V.~K., {et~al.} 1991, Current
  Science, 60, 95

\bibitem[{{The HERA Collaboration} {et~al.}(2022){The HERA Collaboration},
  {Aguirre}, {Alexander}, {Ali}, {Balfour}, {Beardsley}, {Bernardi},
  {Billings}, {Bowman}, {Bradley}, {Bull}, {Burba}, {Carey}, {Carilli},
  {Cheng}, {DeBoer}, {Dexter}, {de Lera Acedo}, {Dibblee-Barkman}, {Dillon},
  {Ely}, {Ewall-Wice}, {Fagnoni}, {Fritz}, {Furlanetto}, {Gale-Sides},
  {Glendenning}, {Gorthi}, {Greig}, {Grobbelaar}, {Halday}, {Hazelton},
  {Hewitt}, {Hickish}, {Jacobs}, {Julius}, {Kern}, {Kerrigan}, {Kittiwisit},
  {Kohn}, {Kolopanis}, {Lanman}, {La Plante}, {Lekalake}, {Lewis}, {Liu},
  {MacMahon}, {Malan}, {Malgas}, {Maree}, {Martinot}, {Matsetela}, {Mesinger},
  {Molewa}, {Morales}, {Mosiane}, {Murray}, {Neben}, {Nikolic}, {Nunhokee},
  {Parsons}, {Patra}, {Pascua}, {Pieterse}, {Pober}, {Razavi-Ghods},
  {Ringuette}, {Robnett}, {Rosie}, {Sims}, {Singh}, {Smith}, {Syce},
  {Thyagarajan}, {Williams}, {Zheng}, \& {HERA
  Collaboration}}]{2022ApJ...925..221A}
{The HERA Collaboration}, Z., {Aguirre}, J.~E., {Alexander}, P., {et~al.} 2022,
  \apj, 925, 221, \dodoi{10.3847/1538-4357/ac1c78}

\bibitem[{{Thyagarajan} {et~al.}(2015){Thyagarajan}, {Jacobs}, {Bowman},
  {Barry}, {Beardsley}, {Bernardi}, {Briggs}, {Cappallo}, {Carroll}, {Corey},
  {de Oliveira-Costa}, {Dillon}, {Emrich}, {Ewall-Wice}, {Feng}, {Goeke},
  {Greenhill}, {Hazelton}, {Hewitt}, {Hurley-Walker}, {Johnston-Hollitt},
  {Kaplan}, {Kasper}, {Kim}, {Kittiwisit}, {Kratzenberg}, {Lenc}, {Line},
  {Loeb}, {Lonsdale}, {Lynch}, {McKinley}, {McWhirter}, {Mitchell}, {Morales},
  {Morgan}, {Neben}, {Oberoi}, {Offringa}, {Ord}, {Paul}, {Pindor}, {Pober},
  {Prabu}, {Procopio}, {Riding}, {Rogers}, {Roshi}, {Udaya Shankar}, {Sethi},
  {Srivani}, {Subrahmanyan}, {Sullivan}, {Tegmark}, {Tingay}, {Trott},
  {Waterson}, {Wayth}, {Webster}, {Whitney}, {Williams}, {Williams}, {Wu}, \&
  {Wyithe}}]{2015ApJ...804...14T}
{Thyagarajan}, N., {Jacobs}, D.~C., {Bowman}, J.~D., {et~al.} 2015, \apj, 804,
  14, \dodoi{10.1088/0004-637X/804/1/14}

\bibitem[{{Tingay} {et~al.}(2013){Tingay}, {Goeke}, {Bowman}, {Emrich}, {Ord},
  {Mitchell}, {Morales}, {Booler}, {Crosse}, {Wayth}, {Lonsdale}, {Tremblay},
  {Pallot}, {Colegate}, {Wicenec}, {Kudryavtseva}, {Arcus}, {Barnes},
  {Bernardi}, {Briggs}, {Burns}, {Bunton}, {Cappallo}, {Corey}, {Deshpande},
  {Desouza}, {Gaensler}, {Greenhill}, {Hall}, {Hazelton}, {Herne}, {Hewitt},
  {Johnston-Hollitt}, {Kaplan}, {Kasper}, {Kincaid}, {Koenig}, {Kratzenberg},
  {Lynch}, {Mckinley}, {Mcwhirter}, {Morgan}, {Oberoi}, {Pathikulangara},
  {Prabu}, {Remillard}, {Rogers}, {Roshi}, {Salah}, {Sault}, {Udaya-Shankar},
  {Schlagenhaufer}, {Srivani}, {Stevens}, {Subrahmanyan}, {Waterson},
  {Webster}, {Whitney}, {Williams}, {Williams}, \& {Wyithe}}]{tingay13}
{Tingay}, S.~J., {Goeke}, R., {Bowman}, J.~D., {et~al.} 2013, \pasa, 30, e007,
  \dodoi{10.1017/pasa.2012.007}

\bibitem[{{Trott} {et~al.}(2016){Trott}, {Pindor}, {Procopio}, {Wayth},
  {Mitchell}, {McKinley}, {Tingay}, {Barry}, {Beardsley}, {Bernardi}, {Bowman},
  {Briggs}, {Cappallo}, {Carroll}, {de Oliveira-Costa}, {Dillon}, {Ewall-Wice},
  {Feng}, {Greenhill}, {Hazelton}, {Hewitt}, {Hurley-Walker},
  {Johnston-Hollitt}, {Jacobs}, {Kaplan}, {Kim}, {Lenc}, {Line}, {Loeb},
  {Lonsdale}, {Morales}, {Morgan}, {Neben}, {Thyagarajan}, {Oberoi},
  {Offringa}, {Ord}, {Paul}, {Pober}, {Prabu}, {Riding}, {Udaya Shankar},
  {Sethi}, {Srivani}, {Subrahmanyan}, {Sullivan}, {Tegmark}, {Webster},
  {Williams}, {Williams}, {Wu}, \& {Wyithe}}]{2016ApJ...818..139T}
{Trott}, C.~M., {Pindor}, B., {Procopio}, P., {et~al.} 2016, \apj, 818, 139,
  \dodoi{10.3847/0004-637X/818/2/139}

\bibitem[{{Trott} {et~al.}(2020){Trott}, {Jordan}, {Midgley}, {Barry}, {Greig},
  {Pindor}, {Cook}, {Sleap}, {Tingay}, {Ung}, {Hancock}, {Williams}, {Bowman},
  {Byrne}, {Chokshi}, {Hazelton}, {Hasegawa}, {Jacobs}, {Joseph}, {Li}, {Line},
  {Lynch}, {McKinley}, {Mitchell}, {Morales}, {Ouchi}, {Pober}, {Rahimi},
  {Takahashi}, {Wayth}, {Webster}, {Wilensky}, {Wyithe}, {Yoshiura}, {Zhang},
  \& {Zheng}}]{trott20}
{Trott}, C.~M., {Jordan}, C.~H., {Midgley}, S., {et~al.} 2020, \mnras, 493,
  4711, \dodoi{10.1093/mnras/staa414}

\bibitem[{{van der Walt} {et~al.}(2011){van der Walt}, {Colbert}, \&
  {Varoquaux}}]{numpy}
{van der Walt}, S., {Colbert}, S.~C., \& {Varoquaux}, G. 2011, Computing in
  Science Engineering, 13, 22

\bibitem[{{van Haarlem} {et~al.}(2013){van Haarlem}, {Wise}, {Gunst}, {Heald},
  {McKean}, {Hessels}, {de Bruyn}, {Nijboer}, {Swinbank}, {Fallows},
  {Brentjens}, {Nelles}, {Beck}, {Falcke}, {Fender}, {H{\"o}randel},
  {Koopmans}, {Mann}, {Miley}, {R{\"o}ttgering}, {Stappers}, {Wijers},
  {Zaroubi}, {van den Akker}, {Alexov}, {Anderson}, {Anderson}, {van Ardenne},
  {Arts}, {Asgekar}, {Avruch}, {Batejat}, {B{\"a}hren}, {Bell}, {Bell}, {van
  Bemmel}, {Bennema}, {Bentum}, {Bernardi}, {Best}, {B{\^\i}rzan}, {Bonafede},
  {Boonstra}, {Braun}, {Bregman}, {Breitling}, {van de Brink}, {Broderick},
  {Broekema}, {Brouw}, {Br{\"u}ggen}, {Butcher}, {van Cappellen}, {Ciardi},
  {Coenen}, {Conway}, {Coolen}, {Corstanje}, {Damstra}, {Davies}, {Deller},
  {Dettmar}, {van Diepen}, {Dijkstra}, {Donker}, {Doorduin}, {Dromer}, {Drost},
  {van Duin}, {Eisl{\"o}ffel}, {van Enst}, {Ferrari}, {Frieswijk}, {Gankema},
  {Garrett}, {de Gasperin}, {Gerbers}, {de Geus}, {Grie{\ss}meier}, {Grit},
  {Gruppen}, {Hamaker}, {Hassall}, {Hoeft}, {Holties}, {Horneffer}, {van der
  Horst}, {van Houwelingen}, {Huijgen}, {Iacobelli}, {Intema}, {Jackson},
  {Jelic}, {de Jong}, {Juette}, {Kant}, {Karastergiou}, {Koers}, {Kollen},
  {Kondratiev}, {Kooistra}, {Koopman}, {Koster}, {Kuniyoshi}, {Kramer},
  {Kuper}, {Lambropoulos}, {Law}, {van Leeuwen}, {Lemaitre}, {Loose}, {Maat},
  {Macario}, {Markoff}, {Masters}, {McFadden}, {McKay-Bukowski}, {Meijering},
  {Meulman}, {Mevius}, {Middelberg}, {Millenaar}, {Miller-Jones}, {Mohan},
  {Mol}, {Morawietz}, {Morganti}, {Mulcahy}, {Mulder}, {Munk}, {Nieuwenhuis},
  {van Nieuwpoort}, {Noordam}, {Norden}, {Noutsos}, {Offringa}, {Olofsson},
  {Omar}, {Orr{\'u}}, {Overeem}, {Paas}, {Pand ey-Pommier}, {Pandey}, {Pizzo},
  {Polatidis}, {Rafferty}, {Rawlings}, {Reich}, {de Reijer}, {Reitsma},
  {Renting}, {Riemers}, {Rol}, {Romein}, {Roosjen}, {Ruiter}, {Scaife}, {van
  der Schaaf}, {Scheers}, {Schellart}, {Schoenmakers}, {Schoonderbeek},
  {Serylak}, {Shulevski}, {Sluman}, {Smirnov}, {Sobey}, {Spreeuw}, {Steinmetz},
  {Sterks}, {Stiepel}, {Stuurwold}, {Tagger}, {Tang}, {Tasse}, {Thomas},
  {Thoudam}, {Toribio}, {van der Tol}, {Usov}, {van Veelen}, {van der Veen},
  {ter Veen}, {Verbiest}, {Vermeulen}, {Vermaas}, {Vocks}, {Vogt}, {de Vos},
  {van der Wal}, {van Weeren}, {Weggemans}, {Weltevrede}, {White}, {Wijnholds},
  {Wilhelmsson}, {Wucknitz}, {Yatawatta}, {Zarka}, {Zensus}, \& {van
  Zwieten}}]{vanhaarlem13}
{van Haarlem}, M.~P., {Wise}, M.~W., {Gunst}, A.~W., {et~al.} 2013, \aap, 556,
  A2, \dodoi{10.1051/0004-6361/201220873}

\bibitem[{{Virtanen} {et~al.}(2020){Virtanen}, {Gommers}, {Oliphant},
  {Haberland}, {Reddy}, {Cournapeau}, {Burovski}, {Peterson}, {Weckesser},
  {Bright}, {van der Walt}, {Brett}, {Wilson}, {Jarrod Millman}, {Mayorov},
  {Nelson}, {Jones}, {Kern}, {Larson}, {Carey}, {Polat}, {Feng}, {Moore}, {Vand
  erPlas}, {Laxalde}, {Perktold}, {Cimrman}, {Henriksen}, {Quintero}, {Harris},
  {Archibald}, {Ribeiro}, {Pedregosa}, {van Mulbregt}, \&
  {Contributors}}]{2020SciPy-NMeth}
{Virtanen}, P., {Gommers}, R., {Oliphant}, T.~E., {et~al.} 2020, Nature
  Methods, 17, 261, \dodoi{https://doi.org/10.1038/s41592-019-0686-2}

\bibitem[{{Wandelt} {et~al.}(2004){Wandelt}, {Larson}, \&
  {Lakshminarayanan}}]{2004PhRvD..70h3511W}
{Wandelt}, B.~D., {Larson}, D.~L., \& {Lakshminarayanan}, A. 2004, \prd, 70,
  083511, \dodoi{10.1103/PhysRevD.70.083511}

\bibitem[{{Wayth} {et~al.}(2018){Wayth}, {Tingay}, {Trott}, {Emrich},
  {Johnston-Hollitt}, {McKinley}, {Gaensler}, {Beardsley}, {Booler}, {Crosse},
  {Franzen}, {Horsley}, {Kaplan}, {Kenney}, {Morales}, {Pallot}, {Sleap},
  {Steele}, {Walker}, {Williams}, {Wu}, {Cairns}, {Filipovic}, {Johnston},
  {Murphy}, {Quinn}, {Staveley-Smith}, {Webster}, \&
  {Wyithe}}]{2018PASA...35...33W}
{Wayth}, R.~B., {Tingay}, S.~J., {Trott}, C.~M., {et~al.} 2018, \pasa, 35,
  e033, \dodoi{10.1017/pasa.2018.37}

\bibitem[{{Wilensky} {et~al.}(2020){Wilensky}, {Barry}, {Morales}, {Hazelton},
  \& {Byrne}}]{2020MNRAS.498..265W}
{Wilensky}, M.~J., {Barry}, N., {Morales}, M.~F., {Hazelton}, B.~J., \&
  {Byrne}, R. 2020, \mnras, 498, 265, \dodoi{10.1093/mnras/staa2442}

\bibitem[{{Wilensky} {et~al.}(2022){Wilensky}, {Hazelton}, \&
  {Morales}}]{2022MNRAS.510.5023W}
{Wilensky}, M.~J., {Hazelton}, B.~J., \& {Morales}, M.~F. 2022, \mnras, 510,
  5023, \dodoi{10.1093/mnras/stab3456}

\bibitem[{Wilson {et~al.}(2018)Wilson, Pettit, \& Ostashev}]{wilson2018}
Wilson, D.~K., Pettit, C.~L., \& Ostashev, V. 2018, in Proceedings of Meetings
  on Acoustics 176ASA, Vol.~35, Acoustical Society of America, 055005

\bibitem[{Zhang(2021)}]{Zhang2021}
Zhang, Z. 2021, Journal of Behavioral Data Science, 1, 119–126,
  \dodoi{10.35566/jbds/v1n2/p2}

\end{thebibliography}

\appendix
\vspace{-1.5em}\section{Wiener filter bias}\label{sec:appA}
Despite the Wiener filter being the maximum a posteriori solution for the combined signal $\m{s}$ component of a data vector $\m{d} = \m{s} + \m{n}$, in general it is a biased estimator of $\m{s}$ when the mean of the signal component differs from zero. 
The expectation value of the Wiener filter solution (Eq.~\ref{wf}) is
\begin{equation}
   \langle \m{s}_{\textrm{wf}} \rangle =  \bigg[ \m{S}^{-1} + \m{N}^{-1} \bigg] ^{-1} \m{N}^{-1}  \m{\bar s}, 
\end{equation}
where we have defined $\m{\bar s} \equiv \langle \m{s} \rangle$, i.e. the true mean of the signal. Thus in general $\langle \m{s}_{\textrm{wf}} \rangle \neq \m{s}$, but the bias vanishes when the mean of the signal $\m{\bar s} = 0$. A separate bias is also incurred on the variance of the Wiener filter, which impacts two-point statistics like the power spectrum. 
Taking the covariance of Eq.~\ref{wf}, with data $\m{d}$ drawn from the appropriate statistical ensemble (i.e. not fixed), we find
\begin{align}
    \textrm{Cov}\big(\m{s}_{\textrm{wf}}\big) &=  \langle \m{s}_{\textrm{wf}}~\m{s}_{\textrm{wf}}^\dag \rangle - \langle \m{s}_{\textrm{wf}} \rangle \langle \m{s}_{\textrm{wf}}^\dag \rangle \nonumber \\
 &= \bigg[ \m{S}^{-1} + \m{N}^{-1} \bigg] ^{-2} \m{N}^{-2} \bigg[\m{S} + \m{N} - \m{\bar s}\m{\bar s}^\dag \bigg],  \label{varwf}
\end{align}
where the noise covariance $\m{N}$ has been assumed to be diagonal. If the data are instead treated as being fixed, we find $\textrm{Cov}\big(\m{s}_{\textrm{wf}}\big) = 0$, which is equivalent to the statement that the Wiener filter with given data and prior assumptions is unique. Comparing with the covariance of the GCR equation solution (Eq.~\ref{cr}), and noting that independent Gaussian random variables are uncorrelated, $\langle \boldsymbol{\omega}_0 \boldsymbol{\omega}_1^\dag \rangle = \langle \boldsymbol{\omega}_0 \m{n}^\dag \rangle = \langle \boldsymbol{\omega}_0 \m{s}^\dag \rangle =0 $ and $\langle \boldsymbol{\omega}_i \boldsymbol{\omega}_i^\dag \rangle = \m{I} $, we obtain
\begin{align}
    \textrm{Cov}\big(\m{s}_{\textrm{cr}}\big) &= \textrm{Cov}\big(\m{s}_{\textrm{wf}}\big) + \bigg[ \m{S}^{-1} + \m{N}^{-1} \bigg] ^{-2} \m{N}^{-1} + \bigg[ \m{S}^{-1} + \m{N}^{-1} \bigg] ^{-2} \m{S}^{-1}  \nonumber \\
 &=  \textrm{Cov}\big(\m{s}_{\textrm{wf}}\big) + \bigg[ \m{S}^{-1} + \m{N}^{-1} \bigg] ^{-1}.   \label{varcr}
\end{align}
GCR equation solutions $\m{s}_{\textrm{cr}}$ therefore have strictly greater variance than Wiener filter solutions $\m{s}_{\textrm{wf}}$, and the additional variance has the form expected from the rearrangement shown in Eq.~\ref{posteriorcr}: a sum-of-inverses, stemming from the product of Gaussian distributions in the posterior of Eq.~\ref{post}. This difference in variance is expected, as the Wiener filter is a summary statistic (the maximum a posteriori estimate); summary statistics can typically be measured with lower variance than the variance of the parent distribution.

\section{Sampling from the complex inverse Wishart distribution}\label{sec:appB}

Eq.~\ref{cinvw} requires us to draw samples from a complex inverse Wishart distribution. This appendix briefly describes a practical algorithm to perform the necessary random draws.

Functions to draw random matrices from the real Wishart distribution are available in numerical libraries such as {\tt scipy}. Real inverse Wishart samples can be drawn by noting that the inverse Wishart distribution for a covariance matrix $C$ is closely related to a Wishart distribution where the covariance is replaced by the precision matrix, $P = C^{-1}$ \citep[e.g.][]{Zhang2021}.

A similar connection persists for the complex inverse Wishart distribution, and so to sample from it we only need to be able to sample from a complex Wishart distribution. Unfortunately, a suitable function is not readily available in most standard numerical libraries. A useful observation is that a complex multivariate Gaussian distribution with $N$ parameters can be written as a real multivariate Gaussian with $2N$ parameters by splitting the complex numbers into their real and imaginary parts. Care must be taken to define the blocks of the resulting covariance matrix correctly \citep[e.g.][]{goodman1963}, as the general complex multivariate Gaussian distribution actually depends on two covariance-like objects. For a vector of complex Gaussian random variates $\bf{z}$, these are the covariance, $\bf{C} \propto \bf{z}\bf{z}^\dagger$ (note the conjugate transpose), and the relation matrix, $\bf{D} \propto \bf{z}\bf{z}^{\rm T}$. In this work we are interested in the `circular' case where $\m{D} = 0$, since the relative magnitude of $\bf{D}$ compared with $\bf{C}$ for our visibility simulations is below the expected level of noise given the number of samples used to estimate the covariance ($<0.5\%$), and furthermore the posterior distributions we later aim to sample from are dependent on the information contained in $\bf{C}$, rather than $\bf{D}$. 

With an appropriate splitting of the complex parameters into their real and imaginary parts, and the covariance matrix replaced with the precision matrix as explained above, it is then possible to use the standard real Wishart sampling function to make random draws from the inverse Wishart distribution for the $2N$ real and imaginary parameters. We generate a $2N\times2N$-size real block matrix of the form 
\begin{equation}
\m{C}_{\textrm{block}} = 
\begin{pmatrix}
\m{C}_{RR}   & \m{C}_{IR} \\
\m{C}_{RI}   & \m{C}_{II}   
\end{pmatrix}
\end{equation}
from the complex covariance matrix we wish to sample from using the relations
\begin{align}
    \m{C}_{RR} = \frac{1}{2} \Re(\m{C}); ~~~~
    \m{C}_{II} = \frac{1}{2} \Re(\m{C}); ~~~~
    \m{C}_{IR} = \frac{1}{2} \Im(\m{C}); ~~~~
    \m{C}_{RI} = -\frac{1}{2} \Im(\m{C}),
\end{align}
where the relation matrix $\m{D}$ has been assumed to be zero. We then use a {\tt scipy} routine to generate a real-valued inverse Wishart sample using the $\m{C}_{\textrm{block}}$ matrix, which carries out a Wishart sample using the precision matrix $\m{C}_{\textrm{block}}^{-1}$. This block-form sample contains a random draw of not only the covariance information $\m{C}$ but also the relation information $\m{D}$, and so it is not consistent with $\m{D} = 0$ in general. Because we are only interested in sampling $\m{C}$, we calculate this quantity directly from the block-form sample using the relation
\begin{equation}
    \m{C} = \m{C}_{RR} + \m{C}_{II} + i(\m{C}_{IR}-\m{C}_{RI}).
\end{equation} 
A similar procedure is described in (e.g.) \citet{wilson2018, Zhang2021}.

\end{document}